\documentclass[12pt]{article}
\usepackage{amsmath}
\usepackage{amssymb}
\usepackage{graphicx}
\usepackage{natbib}
\usepackage{url} 
\usepackage[noend]{algpseudocode}
\usepackage{algorithm}
\usepackage{booktabs}
\usepackage{multirow}
\usepackage{setspace}
\usepackage{comment}
\usepackage{float}

\newcommand{\blind}{0}

\newcommand{\bbeta}{\mbox{\boldmath $\beta$}}

\newcommand{\bgamma}{\mbox{\boldmath $\gamma$}}
\newcommand{\bSigma}{\mbox{\boldmath $\Sigma$}}
\algnewcommand\And{\textbf{and}}

\DeclareMathOperator*{\argmin}{arg\,min}

\begin{document}

\def\spacingset#1{\renewcommand{\baselinestretch}%
{#1}\small\normalsize} \spacingset{1}


\if0\blind
{
  \title{\bf Accelerating Proximal Gradient-type Algorithms using Damped Anderson Acceleration with Restarts and Nesterov Initialization}
  \author{Nicholas C. Henderson
    \hspace{.2cm}\\
    Department of Biostatistics, University of Michigan\\
    and \\
    Ravi Varadhan \\
    Department of Oncology and \\
    Deparment of Biostatistics, Johns Hopkins University
    }
    \date{}
  \maketitle
} \fi

\if1\blind
{
  \bigskip
  \bigskip
  \bigskip
  \begin{center}
    {\LARGE\bf Accelerating Proximal Gradient-type \\[0.25mm] Algorithms using Damped Anderson \\[0.25mm] Acceleration with Restarts and Nesterov \\[0.25mm] Initialization \\[0.25mm]}
\end{center}
  \medskip
} \fi

\bigskip
\begin{abstract}
Despite their frequent slow convergence, proximal gradient schemes are widely used in large-scale optimization tasks due to their tremendous stability, scalability, and ease of computation. In this paper, we develop and investigate a general two-phase scheme for accelerating the convergence of proximal gradient algorithms. By using Nesterov's momentum method in an initialization phase, our procedure delivers fast initial descent that is robust to the choice of starting value. Once iterates are much closer to the solution after the first phase, we utilize a variation of Anderson acceleration to deliver more rapid local convergence in the second phase. Drawing upon restarting schemes developed for Nesterov acceleration, we can readily identify points where it is advantageous to switch from the first to the second phase, which enables use of the procedure without requiring one to specify the number of iterations used in each phase. For the second phase, we adapt and extend a version of Anderson acceleration with algorithm restarts, and we introduce a subsetted version of this procedure that improves performance in problems with substantial sparsity. Through simulation studies involving four representative optimization problems, we show that our proposed algorithm can generate substantial improvements over competing acceleration methods. 
\end{abstract}

\noindent%
{\it Keywords:} algorithm restarts; convergence acceleration; momentum method; first-order methods; multi-secant methods
\vfill

\newpage
\spacingset{1.65}

\section{Introduction}
\label{sec:intro}

Large-scale optimization tasks are ubiquitous in modern machine learning and statistics applications. Because of this, first-order methods have recently attracted renewed interest as such methods are usually stable, have low per-iteration computational costs, and are highly amenable to high-dimensional applications. Despite these advantages, first-order schemes such as gradient descent and proximal gradient descent (PGD) are known to often converge very slowly. To address this drawback of first-order methods, a wide range of different accelerated or fast gradient schemes have been proposed (e.g., \cite{su2016}, \cite{kinga2015}, \cite{lin2018}, and \cite{Levy2018}) with most such schemes aiming to maintain the scalability of the original first-order method while providing more rapid convergence.

To accelerate first-order schemes, Nesterov's momentum method (\cite{Nesterov1983}) and variations thereof have received extensive attention in machine learning contexts (\cite{sutskever2013}). This can be attributed to Nesterov acceleration being highly scalable, straightforward to implement, requiring essentially no additional computational effort when compared to the first-order method, and clear guarantees about rates of convergence (\cite{Beck2017}). Although Nesterov acceleration possesses many advantages, utilizing approaches that draw upon a wider range of available acceleration schemes has the potential to provide much more rapid convergence. Acceleration schemes that could be termed as sequence/vector extrapolation or fixed-point acceleration methods have a long history in the applied mathematics and numerical analysis literature (see e.g., \cite{Brezinski2019}) and include techniques such as Aitken's extrapolation, Shanks transformations, and Steffensen's method. These and similar techniques have been used successfully in the statistical literature for their use in accelerating EM and MM algorithms (e.g., \cite{Louis1982}, \cite{kuroda2023}, \cite{agarwal2024}). Another related extrapolation method which has more recently attracted renewed interest is Anderson acceleration (AA) (\cite{Anderson1965}). In addition to demonstrating impressive speed-ups of fixed-point iterations in a variety of contexts, AA is appealing for accelerating large-scale optimization algorithms due to its rapid local convergence, scalability, low storage requirements, and small per-iteration computational cost required for computing the extrapolations.

The main aim of this paper is to harness the strengths of Nesterov's method and AA with a particular focus on their application to accelerating proximal gradient (PG) algorithms. To this end, we propose a hybrid iterative procedure which begins with a ``Nesterov initialization'' phase and then moves to a more stable variant of AA once Nesterov exhibits signs of slowing down. This is primarily motivated by our observation that, for PG schemes, Nesterov acceleration frequently gives faster descent in early iterations than either AA or other sophisticated extrapolation methods.
However, as the Nesterov-accelerated iterates near the solution, it often begins showing large ``bumps'' or ``waves'' in the objective function, and the speed of convergence near the solution can be unimpressive - particularly if a highly accurate solution is desired. Adding algorithm restarts to Nesterov (e.g., \cite{giselsson2014}, \cite{donogue2015}) is an effective tool that addresses this concern and does frequently improve local convergence speed. Nevertheless, while adding algorithm restarts certainly improves the performance of Nesterov, we believe larger improvements can be gained from using an AA-based acceleration scheme after Nesterov begins to oscillate due to the fast local linear convergence often exhibited by many AA schemes.

Our main strategy is to take advantage of the fast local convergence of AA once the iterates have passed through a Nesterov initialization phase. 
However, early iterations of AA can often be unstable and both stabilization measures and objective function progress checks can substantially improve the performance of AA. Because of this, we utilize a recently developed regularized version of AA called DAAREM \citep{henderson2019} that is both more robust due to damped extrapolations and monotonicity monitoring and is faster due to the inclusion of algorithm restarts. It is worth noting that the Nesterov initialization phase usually generates ``initial'' iterates which are close to the solution and thus, the need for
the stabilization measures of DAAREM may seem less salient. Nevertheless, we have found that, even when using 
a Nesterov initialization phase, using the stabilization techniques of DAAREM is more reliable and as fast or faster than AA, and combining these advantages with Nesterov initialization leads to a robust algorithm which generally requires very few gradient updates to converge.

In addition to having a rapid early descent phase, a chief advantage of using a Nesterov initialization phase is the notable robustness of Nesterov's method. In contrast to many other acceleration schemes, Nesterov acceleration has clear guarantees for global convergence despite the fact that Nesterov uses pre-defined weights in its extrapolation and does not require one to monitor the progress of the objective function. Moreover, in our experience, Nesterov is remarkably robust even in problems that are highly ill-conditioned, and in such cases, Nesterov generally converges to the same fixed point as the associated gradient descent algorithm. This robustness virtually guarantees that iterates started far away from the solution will be close to the solution once the Nesterov initialization phase terminates. Hence, DAAREM will start with a very good ``initial value'' which both ensures that DAAREM will have more robust performance and that DAAREM will begin in a region where the rapid local convergence of DAAREM can take over.

This paper has the following organization. We start in Sections 2 and 3 by describing several examples of proximal gradient algorithms and reviewing the FISTA acceleration scheme -- an adaptation of Nesterov acceleration to handle objective functions having a nonsmooth component. In Section 4, we describe the AA and DAAREM methods for accelerating monotone algorithms, and we also outline our modifications to the control parameters of DAAREM to better handle PG algorithms that have sparse parameter updates. Section 5 describes what we refer to as the NIDAAREM algorithm, where iterates are updated using the Nesterov momentum method until some switching criterion is satisfied. Section 5 also introduces a subsetted version of NIDAAREM which can provide more efficient extrapolation in sparse, high-dimensional problems by discarding irrelevant parameter updates. Section 6 discusses key implementation details for NIDAAREM, and Section 7 details the results of numerical experiments evaluating the performance of NIDAAREM and competing methods on several well-known large-scale optimization examples. We conclude with a brief discussion in Section 8.
\vspace{-0.5cm}

\section{Proximal Gradient Algorithms}
\label{sec:pgd_algorithms}
While many modern applications in machine learning and statistics require optimization of nonsmooth functions, the objective function in many such applications may be decomposed into the sum of a smooth function and a simple, nonsmooth function. 
In this paper, we consider minimization of an objective function $\varphi: \Omega \longrightarrow (-\infty, \infty]$ that can be 
represented as the sum of a smooth function $g$ and a nonsmooth function $h$ as follows 
\vspace{-0.4cm}
\begin{equation}
\varphi( \mathbf{x} ) = g( \mathbf{x} ) + h( \mathbf{x} ),
\label{eq:basic_form}
\end{equation}
where $\Omega$ is a finite-dimensional Euclidean space. 
It is common to assume that $g: \Omega \longrightarrow \mathbb{R}$ is smooth in the sense that it is
$L_{g}$-smooth, which means that $g$ is differentiable on $\Omega$ and there is a Lipschitz constant $L_{g} > 0$ such that 
\begin{equation}
|| \nabla g( \mathbf{x} ) - \nabla g( \mathbf{y} ) || \leq L_{g}|| \mathbf{x} - \mathbf{y} ||, \qquad \textrm{ for all } \mathbf{x},\mathbf{y} \in \Omega. \nonumber
\end{equation}

Proximal gradient (PG) algorithms for minimizing (\ref{eq:basic_form}) proceed by minimizing the sum of $h$ plus a quadratic approximation to $g$. Starting with the $k^{th}$ iterate $\mathbf{x}_{k} \in \Omega$, the PG update of $\mathbf{x}_{k}$ with a steplength of $t$ is given by
\begin{eqnarray}
\mathbf{x}_{k+1} &=& \argmin_{z \in \Omega}\Big\{ g(\mathbf{x}_{k}) + (\mathbf{z} - \mathbf{x})^{T}\nabla g(\mathbf{x}_{k} \rangle + \frac{1}{2t}|| \mathbf{z} - \mathbf{x}_{k}||^{2} + h(\mathbf{z}) \Big\}  \nonumber \\
&=& \textrm{prox}_{th}\big( \mathbf{x}_{k} - t \nabla g( \mathbf{x}_{k} ) \big),
\label{eq:proximal_gradient_update}
\end{eqnarray}
where, in (\ref{eq:proximal_gradient_update}), $\textrm{prox}_{th}: \Omega \longrightarrow \Omega$ refers to the proximal mapping of the function $th(x)$. This mapping is defined as
\begin{equation}
\textrm{prox}_{th}( \mathbf{x} ) = \argmin_{\mathbf{z} \in \Omega}\Big\{ th(\mathbf{z}) + \frac{1}{2}|| \mathbf{z} - \mathbf{x} ||^{2} \Big\}.
\label{eq:prox_mapping}
\end{equation}

The class of optimization problems falling into the class represented by (\ref{eq:basic_form}) is quite broad and includes smooth convex optimization, minimization in sparse regression problems with non-smooth regularizers, and constrained convex optimization. PG algorithms can be used to solve a wide range of optimization tasks commonly encountered in high-dimensional statistics and machine learning (\cite{parikh2014}). Three examples of PG algorithms and how they fit into the notation introduced above are detailed below, and additional examples are studied in Section 6.

\smallskip

\noindent
\textbf{Example 1: $l_{1}$-regularized Regression.}
Suppose we have an $n \times 1$ vector of continuous observations $\mathbf{y} = (Y_{1}, \ldots, Y_{n})^{T}$ and an associated design matrix $\mathbf{X} \in \mathbb{R}^{n \times p}$. 
To perform simultaneous variable selection and shrinkage, it is common to add an $l_{1}$ penalty to the sum-of-squares criterion. In this case, the objective function $\varphi: \Omega \longrightarrow \mathbb{R}$ is 
\begin{equation}
\varphi(\bbeta) = \frac{1}{2}( \mathbf{y} - \mathbf{X}\bbeta)^{T}( \mathbf{y} - \mathbf{X}\bbeta) + \lambda \sum_{j=1}^{p}| \beta_{j}|, \nonumber 
\end{equation}
where $\Omega = \mathbb{R}^{p}$, $g(\bbeta) = \frac{1}{2}|| \mathbf{y} - \mathbf{X}\bbeta ||_{2}^{2}$, and $h(\bbeta) = \lambda\sum_{j=1}^{p} | \beta_{j} |$.
The gradient of $g$ is $\nabla g(\bbeta) = -\mathbf{X}^{T}(\mathbf{y} - \mathbf{X}\bbeta)$, and it can be shown that, with respect to the $l_{2}$ norm, $\nabla g$ has Lipschitz constant $L_{g} = \lambda_{max}( \mathbf{X}^{T}\mathbf{X})$, where $\lambda_{max}( \mathbf{X}^{T}\mathbf{X})$ denotes the maximal eigenvalue of $\mathbf{X}^{T}\mathbf{X}$. 
The PG mapping (\ref{eq:prox_mapping}) in this example can be shown to have the form
\begin{equation}
G_{t}( \bbeta ) = \textrm{prox}_{th}\Big(\bbeta - t \nabla g(\bbeta) \Big)
= S_{t\lambda}\Big( \bbeta + t\mathbf{X}^{T}\{ \mathbf{y} - \mathbf{X}\bbeta \} \Big),
\end{equation}
where $S_{\lambda}: \mathbb{R}^{p} \longrightarrow \mathbb{R}^{p}$ is the ``soft-thresholding'' mapping whose $j^{th}$ component is
$[S_{\lambda}(\bbeta)]_{j} = \textrm{sign}( \beta_{j} )(|\beta_{j}| - \lambda)_{+}$,
with the notation $x_{+}$ meaning $x_{+} = \max\{ x, 0 \}$.

\noindent
\textbf{Example 2: $\ell_{1}$-penalized Logistic Regression.}Consider a vector of binary labels $\mathbf{y} = (Y_{1}, \ldots, Y_{n})^{T}$, i.e.,  $Y_{i} \in \{0, 1\}$ for $i = 1,\ldots,n$ and a design matrix $\mathbf{X} \in \mathbb{R}^{n \times p}$ whose $i^{th}$ row is $\mathbf{x}_{i}^{T}$. Under the assumptions that the outcomes $Y_{i}$ are  independent and that the probability of observing $Y_{i}=1$ is $1/\{1 + \exp( -\mathbf{x}_{i}^{T}\bbeta) \}$, the $\ell_{1}$-penalized negative log-likelihood function is given by
\begin{eqnarray}
\varphi(\bbeta) 
&=& \mathbf{1}_{n}^{T}\log\Big(1 + \exp(\mathbf{X}\bbeta) \Big) - \bbeta^{T}\mathbf{X}^{T}\mathbf{y}  + \lambda\sum_{j=1}^{p}| \beta_{j} |, 
\label{eq:logistic_objective}
\end{eqnarray}
where $\mathbf{1}_{n}$ denotes the $n \times 1$ vector $\mathbf{1}_{n} = (1, \ldots, 1)^{T}$.
In this case, $\Omega = \mathbb{R}^{p}$, and objective function (\ref{eq:logistic_objective}) can be expressed as
\begin{equation}
\varphi(\bbeta) = g(\bbeta) + h(\bbeta), \nonumber 
\end{equation}
where $g( \bbeta) = \mathbf{1}_{n}^{T}\log\Big(1 + \exp(\mathbf{X}\bbeta) \Big) - \bbeta^{T}\mathbf{X}^{T}\mathbf{y}$, and $h( \bbeta ) = \lambda\sum_{j=1}^{p} | \beta_{j} |$. The gradient is $\nabla g(\bbeta) = \mathbf{X}^{T}(\theta( \mathbf{X}\bbeta) - \mathbf{y})$, where $\theta( \mathbf{X}\bbeta )$ denotes the $n \times 1$ vector whose $i^{th}$ component is $1/\{1 + \exp( -\mathbf{x}_{i}^{T}\bbeta) \}$. It can be shown that the matrix $\nabla^{2} g(\bbeta) - \tfrac{1}{4}\mathbf{X}^{T}\mathbf{X}$ is negative definite for any $\bbeta$, and hence  
$\nabla g$ has Lipschitz constant $L_{g} = \lambda_{max}( \mathbf{X}^{T}\mathbf{X})/4$. The proximal gradient mapping in this example is
\begin{eqnarray}
G_{t}( \bbeta ) &=& \textrm{prox}_{th}\Big(\bbeta - t \nabla g(\bbeta) \Big)
= \textrm{prox}_{th}\Big(\bbeta - t  \mathbf{X}^{T}(\theta( \mathbf{X}\bbeta) - \mathbf{y}) \Big) \nonumber \\
&=& \argmin_{\boldsymbol{z}} \Bigg[ t\lambda\sum_{j=1}^{p} | z_{j}| + \frac{1}{2}\{ \mathbf{z} - \bbeta + t  \mathbf{X}^{T}(\theta( \mathbf{X}\bbeta) - \mathbf{y})\}^{T}\{ \mathbf{z} - \bbeta + t  \mathbf{X}^{T}(\theta( \mathbf{X}\bbeta) - \mathbf{y})\} \Bigg] \nonumber \\
&=& S_{t\lambda}\Big( \bbeta + t\mathbf{X}^{T}\{ \mathbf{y} - \theta( \mathbf{X}\bbeta) \} \Big), \nonumber 
\end{eqnarray}
where $S_{\lambda}: \mathbb{R}^{p} \longrightarrow \mathbb{R}^{p}$ is the soft-thresholding mapping. 

\noindent
\textbf{Example 3: Matrix Completion.}
\label{sec:matrix_complete}
The aim of matrix completion (\cite{Candes2009}) is to fill in the missing entries of an $n \times p$  matrix $\mathbf{A}$ whose entries are only observed for some subset of indices $\Theta$. Specifically, one observes the $(i,j)$ component of $\mathbf{A}$ if $(i,j) \in \Theta$ while the $(i,j)$ component of $\mathbf{A}$ is missing if $(i,j) \not\in \Theta$. If one defines the projection $P_{\Theta}( \mathbf{Z} ) \in \mathbb{R}^{n \times p}$ of the matrix $\mathbf{Z} \in \mathbb{R}^{n \times p}$ onto the set of observed indices 
\begin{equation}
[P_{\Theta}( \mathbf{Z} )]_{ij}
= \begin{cases}
Z_{ij} & \text{ if } (i,j) \in \Theta \\
0 & \text{ otherwise,} \nonumber
\end{cases}
\end{equation}
then the objective function $\varphi:\mathbb{R}^{n \times p} \longrightarrow \mathbb{R}$ in matrix completion is
\begin{eqnarray}
\varphi( \mathbf{Z} ) &=&  \frac{1}{2}\sum_{(i,j)\in\Theta}( Z_{ij} - A_{ij})^{2} + \lambda|| \mathbf{Z} ||_{*}
\nonumber \\
&=& \frac{1}{2}|| P_{\Theta}( \mathbf{A} ) - P_{\Theta}(\mathbf{Z}) ||_{F}^{2} + \lambda|| \mathbf{Z} ||_{*}, \nonumber 
\end{eqnarray}
where $|| \mathbf{B} ||_{F} = \sqrt{\textbf{tr}( \mathbf{B}^{T}\mathbf{B})}$ is the Frobenius norm and $|| \mathbf{B} ||_{*} = \sum_{k=1}^{\min\{ n, p\}} \sigma_{k}(\mathbf{A})$ (for singular values $\sigma_{k}(\mathbf{A})$ of $\mathbf{A}$) denotes the nuclear norm of $\mathbf{A}$.
In this example, $\Omega = \mathbb{R}^{n \times p}$, $g( \mathbf{Z} ) = \frac{1}{2}|| P_{\Theta}( \mathbf{A} ) - P_{\Theta}(\mathbf{Z}) ||_{F}^{2}$, and $h( \mathbf{Z} ) = \lambda|| \mathbf{Z} ||_{*}$. The gradient of $g$ is $\nabla g(\mathbf{Z}) = P_{\Theta}(\mathbf{Z}) - P_{\Theta}(\mathbf{A})$, and with respect to the Frobenius norm, $\nabla g$ has Lipschitz constant $L_{g} = 1$. 
The proximal mapping in this case can be shown (see, e.g., \cite{Mazumder2010}) to have the form
\begin{eqnarray}
G_{t}( \mathbf{Z} ) &=& \textrm{prox}_{th}\Big(\mathbf{Z} - t \nabla g(\mathbf{Z}) \Big) = \argmin_{\mathbf{B} \in \mathbb{R}^{n \times p}} \frac{1}{2}||\mathbf{B} - \mathbf{Z} + t\{ P_{\Theta}(\mathbf{Z}) - P_{\Theta}( \mathbf{A}) \} ||_{F}^{2} + t\lambda|| \mathbf{B} ||_{*} \nonumber \\
&=& D_{\lambda t}\Big( \mathbf{Z} -  t\{ P_{\Theta}(\mathbf{Z})  - P_{\Theta}( \mathbf{A} ) \} \Big),
\label{eq:matrix_complete_update}
\end{eqnarray}
where $D_{\lambda}( \mathbf{Z} )$ is the SVD soft-thresholding function defined as
$D_{\lambda}( \mathbf{Z} ) = \mathbf{U}_{\mathbf{Z}}\bSigma_{\mathbf{Z},\lambda} \mathbf{V}_{\mathbf{Z}}^{T}$.
Here, $\mathbf{Z} = \mathbf{U}_{\mathbf{Z}}\bSigma_{\mathbf{Z}}\mathbf{V}_{\mathbf{Z}}^{T}$ is the singular value decomposition of $\mathbf{Z}$, $\bSigma_{\mathbf{Z}} \in \mathbb{R}^{p \times p}$ is the diagonal matrix with diagonal entries $d_{1}, \ldots, d_{\min\{n,p\}}$, and $\bSigma_{\mathbf{Z},\lambda} = \textrm{diag}\{ (d_{1} - \lambda)_{+}, \ldots, (d_{\min\{n,p\}} - \lambda)_{+})\}$. 
If the steplength in (\ref{eq:matrix_complete_update}) is set to $t = 1$, then (\ref{eq:matrix_complete_update}) reduces to the Soft-Impute algorithm described in \cite{Mazumder2010}.
\vspace{-0.5cm}

\section{Nesterov Acceleration and FISTA}

\cite{Nesterov1983} proposed a straightforward and intriguing method for speeding the convergence of gradient descent when the objective function under consideration is smooth. Nesterov's method works by updating the current iterate by taking a gradient descent step after adding a ``momentum term'' whose value depends, in part, on the previous iteration. 
\cite{Beck2009} described a reformulation of Nesterov's method to handle nonsmooth composite objective functions having the form (\ref{eq:basic_form}). Due to the fact that many PG updates involve shrinkage and thresholding, \cite{Beck2009} label their method the ``Fast Iterative Shrinkage-Thresholding Algorithm'' or FISTA.  
Nesterov's method and variations thereof such as FISTA are widely used to accelerate the convergence of gradient or PG algorithms. 
\begin{algorithm}[ht]
\caption{(Nesterov acceleration for proximal gradient descent).}\label{euclid}
\begin{algorithmic}[1] \onehalfspacing
\State Given $\mathbf{x}_{0} \in \Omega$, steplength $t \in (0, 1/L_{g})$, and proximal gradient mapping $G_{t}$.
\vskip-1pt
\State Set $\mathbf{y}_{1} = \mathbf{x}_{0}$ and $a_{1} = 1$.
\For{k=1,2,3,...until convergence}
\State $\mathbf{x}_{k} = G_{t}( \mathbf{y}_{k} )$
\State $a_{k+1} = \big\{1 + \sqrt{1 + 4a_{k}^{2}} \big\}/2$
\State $\mathbf{y}_{k + 1} = \mathbf{x}_{k} + \Big( \frac{ a_{k} - 1}{a_{k+1}} \Big)(\mathbf{x}_{k} - \mathbf{x}_{k-1})$.
\EndFor
\end{algorithmic}
\label{alg:fista}
\end{algorithm}

The FISTA scheme with fixed steplength $t$ is described in Algorithm \ref{alg:fista}, and we will
henceforth refer to the procedure detailed in Algorithm \ref{alg:fista} as simply Nesterov's method.
A main advantage of Nesterov's method for accelerating PG schemes is that its implementation is no more complex than an implementation of PGD itself, as it only requires performing a single PG step within each iteration. Despite its remarkable simplicity, Nesterov's method can often deliver substantial gains in convergence speed when compared with PGD. 
\vspace{-0.5cm}

\section{Convergence Acceleration of Proximal Gradient Algorithms via DAAREM and Related Schemes}
\vspace{-0.3cm}

\subsection{Anderson Acceleration and Algorithm Restarts}
While acceleration schemes based on Nesterov's method have been widely used to speed convergence of first-order methods, there are a number of alternative acceleration methods that have the potential to substantially improve upon Nesterov. One attractive class of scalable and efficient acceleration schemes is the Anderson acceleration (AA) (\cite{Anderson1965} and \cite{Walker2011}) scheme that was developed for accelerating fixed point iterations. For a given order $m_{k} \geq 1$ of the AA scheme that can vary with $k$, the iterate $\mathbf{x}_{k+1}$ in the $(k+1)^{st}$ iteration is found by using information from the current and past $m_{k}$ iterations. Specifically, the AA update $\mathbf{x}_{k+1}$ is obtained as a linear combination of the proximal gradient mappings $G_{t}(\mathbf{x}_{k})$ of the most recent $m_{k}$ iterates
\begin{equation}
\mathbf{x}_{k+1} = G_{t}(\mathbf{x}_{k}) - \sum_{j=1}^{m_{k}} \gamma_{j}^{(k)} \Big[ G_{t}( \mathbf{x}_{k + j - m_{k}} )
- G_{t}( \mathbf{x}_{k + j - 1 - m_{k}} ) \Big],  
\label{eq:basic_aa_update}
\end{equation}
where the $m_{k} \times 1$ vector of AA coefficients $\bgamma^{(k)} = (\gamma_{1}^{(k)}, \ldots, \gamma_{m_{k}}^{(k)})^{T}$ is found by solving the following least-squares problem
\begin{equation}
\bgamma^{(k)} = \argmin_{\bgamma} || \mathbf{f}_{k} - \mathbf{F}_{k}\bgamma ||_{2}. 
\label{eq:aa_coefficients}
\end{equation}
In (\ref{eq:aa_coefficients}), $\mathbf{f}_{k} \in \mathbb{R}^{p}$ denotes the residual vector $\mathbf{f}_{k} = G_{t}(\mathbf{x}_{k}) - \mathbf{x}_{k}$ and $\mathbf{F}_{k} \in \mathbb{R}^{p \times m_{k}}$ denotes the matrix whose $j^{th}$ column is $G_{t}(\mathbf{x}_{k + j - m_{k} - 1}) - \mathbf{x}_{k + j - m_{k} -1}$.

While requiring more computational effort per iteration than Nesterov's method, the additional per-iteration computational requirements for AA are modest even in very high-dimensional problems. The main additional per-iteration requirement in AA is solving the least squares problem $\min_{\bgamma} || \mathbf{f}_{k} - \mathbf{F}_{k} \bgamma ||_{2}$, where $\mathbf{f}_{k} \in \mathbb{R}^{p}$ and $\mathbf{F}_{k} \in \mathbb{R}^{p \times m_{k}}$. Because $m_{k}$ is usually quite small, this least-squares problem is manageable even for very large values of $p$. 
Moreover, the additional storage requirements in AA are quite modest as one only needs to store the two long but ``thin'' $p \times m_{k}$ matrices $\mathbf{X}_{k}$ and $\mathbf{F}_{k}$.

In many settings, schemes for accelerating fixed-point iterations can benefit from periodic {\it algorithm restarts}. Here, restarts refers to periodically treating an iterate as if it were the initial value and starting the acceleration scheme anew from that point. The advantages of including restarts have been noted in a  
variety of other acceleration methods. For example, in the context of accelerating general fixed-point iterations, \cite{Smith1987} suggest restarting by intermittently performing a single fixed-point iteration. For conjugate gradient methods, it is often beneficial to periodically take a simple gradient step (e.g., \cite{nocedal2006}). Restart has also been effectively used in the SQUAREM family of schemes for accelerating fixed-point iterations (see \cite{Varadhan2008}). Restarts are often used to improve the performance of the generalized minimal residual method (GMRES) (\cite{Saad1986}) for solving large, nonsymmetric systems of linear equations, and restarting schemes have also been noted to substantially improve the performance of Nesterov acceleration (\cite{donogue2015} and \cite{roulet2017}).

Restart strategies can be either systematic, adaptive, or some combination of these two. With a systematic restart strategy, one restarts the acceleration scheme after every $s_{k}$ iterations, where $s_{k}$ is a pre-specified sequence that does not depend on the progress of the objective function. In the context of AA where the maximal order $m$ is specified in advance, an appealing systematic restart strategy is to restart the acceleration scheme every $m$ iterations. In contrast to systematic approaches, adaptive restart strategies restart whenever a specified condition is violated; for example, one might restart the algorithm whenever a monotonicity condition for the objective function is violated. 
In the context of Nesterov acceleration of PGD, \cite{donogue2015} found improved performance for both fixed and adaptive restarting schemes with the adaptive restarts based on either a monotonicity or a gradient condition violation.
\cite{giselsson2014} also report improved performance when using a different non-monotonicity implying test for restarting an accelerated PG scheme. In the context of AA, (\cite{Pratapa2015} and \cite{henderson2019}) employed systematic restarts, \cite{Fang2009} proposed an adaptive restart strategy, and \cite{zhang2020} developed a restart procedure which is a hybrid of systematic and adaptive approaches. 
Algorithm \ref{alg:restarted_aa} describes a modified version of AA where a systematic restart strategy is used.
In Algorithm \ref{alg:restarted_aa}, AA is restarted every $m$ iterations, 
and in the remainder of the paper, we are referring to the restart scheme described in Algorithm \ref{alg:restarted_aa} whenever we refer to an AA algorithm that uses restarts. 


\begin{algorithm}[ht]
\caption{(Anderson Acceleration with Systematic Restarts).
In the description of the algorithm,
$f_{t}(\mathbf{x}) = G_{t}(\mathbf{x}) - \mathbf{x}$, $\Delta x_{i} = \mathbf{x}_{i+1} - \mathbf{x}_{i}$, $\mathbf{f}_{i} = f_{t}( \mathbf{x}_{i} )$, $\Delta \mathbf{f}_{i} = \mathbf{f}_{i+1} - \mathbf{f}_{i}$, $\mathbf{X}_{k} \in \mathbb{R}^{p \times m_{k}}$ denotes the matrix 
$\mathbf{X}_{k} = \big[ \Delta \mathbf{x}_{k - m_{k}}, \ldots, \Delta \mathbf{x}_{k-1} \big]$, and $\mathbf{F}_{k} \in \mathbb{R}^{p \times m_{k}}$ denotes the matrix
$\mathbf{F}_{k} = 
\big[ \Delta \mathbf{f}_{k-m_{k}}, \ldots, \Delta \mathbf{f}_{k-1} \big]$.
}\label{euclid}
\begin{algorithmic}[1] \onehalfspacing
\State Given $\mathbf{x}_{0} \in \Omega$ and an integer $m \geq 1$.
\State Set $c_{1} = 1$; $\mathbf{x}_{1} = \mathbf{x}_{0} + f_{t}( \mathbf{x}_{0} )$.
\For{k=1,2,3,...until convergence}
\State Set $m_{k} = \min(m, c_{k})$ and compute $\mathbf{f}_{k} = f_{t}(\mathbf{x}_{k})$.
\State Find $\bgamma^{(k)} \in \mathbb{R}^{m_{k}}$ such that
$\bgamma^{(k)} = \argmin_{\bgamma \in \mathbb{R}^{m_{k}}} || \mathbf{f}_{k} - \mathbf{F}_{k}\bgamma||_{2}^{2}$
\State $\mathbf{x}_{k+1} = \mathbf{x}_{k} + \mathbf{f}_{k} - (\mathbf{X}_{k} + \mathbf{F}_{k})\bgamma^{(k)}$
\If {$k$ mod $m = 0$,} set $c_{k+1} = 1$
\Else { set $c_{k+1} = c_{k} + 1$}
\EndIf
\EndFor
\end{algorithmic}
\label{alg:restarted_aa}
\end{algorithm}

\vspace{-0.3cm}

\subsection{DAAREM: Regularizing Anderson Steps and Monotonicity Monitoring} \label{ss:daarem}
While AA is often effective at accelerating convergence, AA can be unstable, particularly in early iterations where the iterates are far away from the solution, and running AA without any additional safeguards or stabilization measures can jeopardize the performance of the acceleration scheme. One way to stabilize the AA steps is to ``dampen'' the coefficients used in the AA extrapolation by solving a penalized least squares problem. This is done in the 
Damped Anderson Acceleration with Restarts and Epsilon Monotonicity (DAAREM) algorithm (\cite{henderson2019}),
where the coefficients used in the AA updates are found by solving the following penalized version of the AA least squares equation
\begin{equation}
(\mathbf{F}_{k}^{T}\mathbf{F}_{k} + \lambda_{k}\mathbf{I}_{m_{k}})\bgamma^{(k)} = \mathbf{F}_{k}^{T}\mathbf{f}_{k}^{T}.
\label{eq:penalized_ls_AA}
\end{equation}
In (\ref{eq:penalized_ls_AA}), the terms $\mathbf{F}_{k}$ and $\mathbf{f}_{k}$ are as defined in Algorithm \ref{alg:restarted_aa},
and $\lambda_{k} \geq 0$ is a nonnegative ``damping'' parameter.
Using $\bgamma^{(k)}$ in (\ref{eq:penalized_ls_AA}) with a larger
value of $\lambda_{k}$ in early iterations helps to stabilize the initial steps by ensuring the initial steps are much closer to the PG step. 

An additional feature of DAAREM which helps to further stabilize AA steps and to improve acceleration performance is the use of monotonicity control. While adding damping often improves the stability and robustness of AA, AA with damping still sacrifices the monotonicity and global convergence guarantees of the original PGD scheme. Moreover, while Nesterov's method comes with certain global convergence guarantees under fairly broad conditions (\cite{Beck2009}), the convergence guarantees for unmodified AA schemes are more sporadic and less general \citep{toth2015}, although a number of recent results have led to more general results about convergence rates of AA-type schemes (e.g., \cite{scieur2024}). Hence, when using an AA-based algorithm, enforcing some level of monotonicity control is needed for a more robust acceleration algorithm. In our experience, monitoring the monotonicity of DAAREM leads to substantially improved robustness without sacrificing much speed or, in many cases, even improving speed of convergence modestly. Because the PG update is already monotone, altering the acceleration scheme to be monotone is straightforward as one simply takes the PG step whenever a monotonicity condition for a proposed AA-extrapolated iterate is violated. Moreover, because the PG update is already performed as part of the AA extrapolation, performing a monotonicity check only requires an additional evaluation of the objective function and does not require taking additional PGD steps.

\begin{algorithm}[ht]
\caption{DAAREM (Damped Anderson Acceleration with Restarts and Monotonicity Monitoring) for accelerating proximal gradient algorithms. 
In the algorithm description, the terms $f_{t}( \mathbf{x} )$, $\Delta \mathbf{x}_{i}$, $\Delta \mathbf{f}_{i}$, $\mathbf{X}_{k}$, and $\mathbf{F}_{k}$ are as defined in Algorithm \ref{alg:restarted_aa}.}\label{euclid}
\begin{algorithmic}[1] \onehalfspacing
\State Given $\mathbf{x}_{0} \in \Omega$, $\varepsilon > 0$, $\varepsilon_{c} > 0$, $\ldots$, $\rho^{*} > 0$, $D > 0$  and an integer $m \geq 1$.
\State Set $c_{1} = 1$; $s_{1} = 0$; $\mathbf{x}_{1} = \mathbf{x}_{0} + f( \mathbf{x}_{0} )$; $\ell^{*} = \ell( \mathbf{x}_{1} )$.
\For{k=1,2,3,...until convergence}
\State Set $m_{k} = \min(m, c_{k})$, and compute $\mathbf{f}_{k} = f_{t}(x_{k})$
\State Compute the condition number $\rho_{k}$ of $\mathbf{F}_{k}$. \textbf{If} $\rho_{k} > \rho^{*}$, set $s_{k} = \max\{ s_{k} - 1, -D\}$.
\State Update $\delta_{k} = 1/(1 + \alpha^{\kappa - s_{k}})$.
\State Find $\lambda_{k}^{*} \geq 0$ such that $||\tilde{\beta}_{r}(\lambda_{k})||_{2}^{2} = \delta_{k}||\tilde{\beta}_{LS}||_{2}^{2}$, where
\vspace{-.2cm}
\begin{equation}
\tilde{\beta}_{r}(\lambda_{k}) = (\mathbf{F}_{k}^{T}\mathbf{F}_{k} + \lambda_{k}\mathbf{I}_{m_{k}})^{-1}\mathbf{F}_{k}^{T}\mathbf{f}_{k}
\qquad \textrm{and} \qquad
\tilde{\beta}_{LS} = (\mathbf{F}_{k}^{T}\mathbf{F}_{k})^{-1}\mathbf{F}_{k}^{T}\mathbf{f}_{k}.
\nonumber 
\vspace{-.2cm}
\end{equation}
\State For the value of $\lambda_{k}^{*}$ found in the previous step, set $\bgamma^{(k)} = \tilde{\beta}_{r}(\lambda_{k}^{*})$.
\State $\mathbf{y}_{k+1} = \mathbf{x}_{k} + \mathbf{f}_{k} - (\mathbf{X}_{k} + \mathbf{F}_{k})\bgamma^{(k)}$
\If {Accept$(\mathbf{y}_{k+1}, \mathbf{x}_{k}, \varepsilon_{k}) = \textrm{TRUE}$,} 
\State $\mathbf{x}_{k + 1} = \mathbf{y}_{k+1}$; $s_{new} = s_{k} + 1$
\Else
\State $\mathbf{x}_{k+1} = \mathbf{x}_{k} + \mathbf{f}_{k}$; $s_{new} = s_{k}$
\EndIf
\If {$k$ mod $m = 0$ \textbf{ and } CycleAccept$(\varphi(\mathbf{x}_{k+1}), \varphi^{*}, \varepsilon_{c}) = $ TRUE,}
\State $c_{k+1} = 1$; $\varphi^{*} = \varphi(\mathbf{x}_{k+1})$
\ElsIf {$k$ mod $m =0$ \textbf{ and } CycleAccept$(\varphi(\mathbf{x}_{k+1}), \varphi^{*}, \varepsilon_{c}) = $ FALSE,}
\State $c_{k+1} = 1$, $s_{new} = \max\{ s_{new} - m, -D\}$; $\varphi^{*} = \varphi( \mathbf{x}_{k+1})$
\Else
\State $c_{k+1} = c_{k} + 1$
\EndIf
\State $s_{k+1} = s_{new}$
\EndFor
\end{algorithmic}
\label{alg:daarem_alg}
\end{algorithm}

Algorithm \ref{alg:daarem_alg} describes our modified version of the original DAAREM algorithm. 
Algorithm \ref{alg:daarem_alg} is nearly the same as the original DAAREM algorithm described in \cite{henderson2019}
except that we have added an extra step checking the condition number of $\mathbf{F}_{k}$ when determining
the level of damping to apply. We also consider a more general acceptance criterion in line 10 of Algorithm \ref{alg:daarem_alg}.
It is the acceptance condition in line 10 of Algorithm \ref{alg:daarem_alg} that determines the monotonicity 
monitoring strategy, and we consider two separate monotonicity strategies here rather than the single monotonicity monitoring strategy
proposed in \cite{henderson2019}. Both of the two acceptance criteria we consider have the form
\begin{equation}
\textrm{Accept}( \mathbf{y}_{k+1}, \mathbf{x}_{k}, \varepsilon_{k})    = \begin{cases}
\textrm{TRUE} & \textrm{ if } \varphi(\mathbf{y}_{k+1}) \leq \varphi(\mathbf{x}_{k}) + \varepsilon_{k} \\
\textrm{FALSE} & \textrm{ if } \varphi(\mathbf{y}_{k+1}) > \varphi(\mathbf{x}_{k}) + \varepsilon_{k},
\end{cases}
\label{eq:eps_mon_cond}
\end{equation}
and the difference between the two monotonicity monitoring schemes is the choice of $\varepsilon_{k}$.

In the first monotonicity monitoring scheme, $\varepsilon_{k} = \varepsilon$ for all iterations $k$, and we refer to this as the
\textbf{fixed monotonicity control} strategy.
Condition (\ref{eq:eps_mon_cond}) states that the proposed iterate $\mathbf{y}_{k + 1}$ will be accepted provided that $\mathbf{y}_{k+1}$ does not increase the objective function by more than $\varepsilon_{k} > 0$, and setting $\varepsilon_{k} = 0$  for all iterations $k$ will lead to a monotone acceleration scheme.
While setting $\varepsilon_{k} = 0$ leads to a robust acceleration scheme that works well in some settings, our experiences suggest that allowing a moderate amount of non-monotonicity leads to much faster convergence in many problems. The default value for DAAREM suggested by \cite{henderson2019} was $\varepsilon = 0.01$ in the context of EM algorithms, but we use the less stringent $\varepsilon = 1$ in all of our simulation studies.

While choosing $\varepsilon_{k}$ to be a small, positive number $\varepsilon > 0$ for all iterations works well for many problems, we have encountered certain problems where setting $\varepsilon_{k} = 0$ works much better and prevents algorithm stagnation. Because it is difficult to determine beforehand which strategy for choosing $\varepsilon_{k}$ is superior, an
effective alternative strategy is to vary $\varepsilon_{k}$ across iterations. We vary the monotonicity parameter,
by changing it after the end of every ``cycle'' of $m$ iterations. Specifically, $\varepsilon_{k}$ is updated from $\varepsilon_{k-1}$ using the following rule:
\begin{equation}
\varepsilon_{k} =
\begin{cases}
\varepsilon_{k-1} & \textrm{ if } k\mod m \neq 0 \\
0 & \textrm{ if } k \mod m = 0 \textrm{ and } \varepsilon_{k-1} = \varepsilon > 0 \\
\varepsilon & \textrm{ if } k \mod m = 0 \textrm{ and } \varepsilon_{k-1} = 0. 
\end{cases}
\label{eq:alternate_strategy}
\end{equation}
In (\ref{eq:alternate_strategy}), the 
monotonicity parameter $\varepsilon_{k}$ has two possible values and stays fixed within each cycle but
is changed to its alternate value at the end of each cycle. Because of the alternating nature of $\varepsilon_{k}$,
we refer to this as the \textbf{alternating monotonicity control} strategy.

For problems where not enforcing strict monotonicity is beneficial, our experience suggests that $\varepsilon = 0.01$ is too strict, particularly for PG algorithms having non-smooth parameter updates, and we have found the more lax criterion $\varepsilon = 1$ tends to perform better in such situations. Hence, our default implementation of alternating monotonicity control cycles between $\varepsilon_{k} = 1$ and $\varepsilon_{k} = 0$.
Though we have found that this strategy works well in a range of problems, when using acceptance criterion (\ref{eq:eps_mon_cond}) and alternating monotonicity strategy (\ref{eq:alternate_strategy}) it may be worth testing different values of $\varepsilon$, especially in problems where the objective function is scaled much differently than the examples shown in this paper.

The monotonicity-monitoring condition (\ref{eq:eps_mon_cond}) relies on evaluating the objective function in each iteration. In many cases, such an evaluation is relatively cheap and entails modest additional computational effort. To address cases where an additional evaluation of the objective function is deemed too expensive, we also explore the use of a residual-based acceptance criterion. Here, the residual of the $k^{th}$ iterate $\mathbf{x}_{k}$ is defined as $r(\mathbf{x}_{k}) = G_{t}( \mathbf{x}_{k} ) - \mathbf{x}_{k}$,
and the acceptance of a proposed iterate $\mathbf{y}_{k+1}$ is determined by whether or not the norm of the residual
$r(\mathbf{y}_{k+1})$ is sufficiently small compared to the norm of the residual $r(\mathbf{x}_{k})$ of the current iterate. In our simulation studies, we have found that residual-based monitoring performs well in many settings but is notably less robust than direct monitoring of the objective function (see Appendix C).   
Further details about the residual-based acceptance criterion and the cycle-level acceptance criteria used in lines 14 and 16 of Algorithm \ref{alg:daarem_alg} are provided in Appendix A.

\vspace{-0.5cm}

\section{Nesterov Initialization, NIDAAREM, and Subsetted Anderson Acceleration}
\vspace{-0.3cm}
\subsection{Nesterov Initialization and NIDAAREM}
\begin{figure}
\begin{center}
\includegraphics[width=6.5in, height=4.0in]{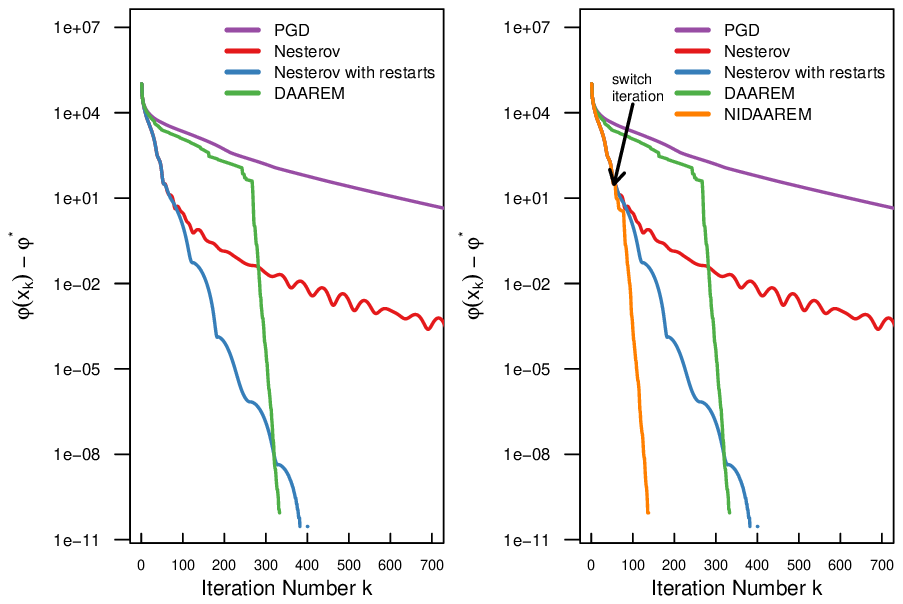}
\end{center}
\vspace{-0.9cm}
\caption{Convergence comparison of Nesterov acceleration, Nesterov with algorithm restarts, DAAREM, and NIDAAREM in an $\ell_{1}$-penalized regression problem. The y axis displays the values of $\varphi(\mathbf{x}_{k}) - \varphi^{*}$,
where $\varphi^{*}$ is the minimum value of the objective function.\label{fig:trace_plots}}
\end{figure}

While Nesterov acceleration usually converges slower than DAAREM, in our experience, early iterations in Nesterov frequently exhibit better progress than the early iterates of DAAREM. This is especially the case for PG algorithms that involve sparse parameter updates. 
An example of this is shown in Figure \ref{fig:trace_plots}, which traces the value of the objective function for different acceleration
methods applied to an $\ell_{1}$-penalized regression problem.
As illustrated in the left-hand panel of Figure \ref{fig:trace_plots}, Nesterov has fast initial progress with rapid objective function declines in the early iterations. Indeed, the objective function for Nesterov is much smaller than any of the competing methods for at least the first 100 iterations. However, Nesterov tends to slow down relative to DAAREM in later iterations, particularly after Nesterov begins exhibiting non-monotonicity and wave-like oscillations. 

The behaviors revealed in the left-hand panel of Figure \ref{fig:trace_plots} suggest that a potentially more efficient strategy is to start off by running Nesterov for a number of iterations and then later switching to DAAREM. This would take advantage of our observation that Nesterov often makes strong progress initially while DAAREM provides very rapid convergence near the solution. However, there are two key concerns in adopting such a ``switching'' approach. First, one would need to construct a criterion to determine exactly where the switch from Nesterov to DAAREM will be made. The switch should ideally occur around the point at which Nesterov begins to slow down from its initial fast descent, but determining a formal criterion for detecting such a slowdown point could be challenging. In this paper, we consider a straightforward monotonicity-based switching criterion which we have found to work well in practice, but it is certainly possible there could be other switching criteria which notably improve performance.
A second concern is that, even though DAAREM would start at a better initial value by using a Nesterov initialization phase, the number of iterations required to converge could be no faster than the usual DAAREM scheme. The concern here is that DAAREM could still exhibit a long ``plateau'' phase which would cause the helpful Nesterov initialization to be mostly useless. 
In all our simulation studies, this has not been the case as DAAREM usually converges rapidly after switching from Nesterov, and the Nesterov initialized DAAREM scheme does not usually have the early ``plateau'' phase exhibited by DAAREM in Figure \ref{fig:trace_plots}.

Algorithm \ref{alg:nidaarem} states our procedure for combining a Nesterov initialization phase with an application of DAAREM until convergence. Line $5$ of this algorithm determines when the Nesterov initialization phase is terminated and the DAAREM iterations begin. This occurs whenever a switching criterion is satisfied or a maximum number $N_{s}$ of Nesterov iterations has occurred. 
To determine when to switch from Nesterov to DAAREM, we have considered the following two criteria proposed by \cite{donogue2015} for Nesterov with restarts
\begin{itemize} \itemsep2pt
    \item
    \textbf{Monotonicity-based Switching Criterion.}
    \vspace{-12pt}
    \begin{center}
        Switch to DAAREM whenever: $\quad \varphi(\mathbf{x}_{k}) > \varphi( \mathbf{x}_{k-1})$. 
    \end{center}
    \item
    \textbf{Gradient-based Switching Criterion.}
    \vspace{-12pt}
    \begin{center}
        Switch to DAAREM whenever: $\quad (\mathbf{y}_{k} - \mathbf{x}_{k})^{T}(\mathbf{x}_{k} - \mathbf{x}_{k-1}) > 0$,
    \end{center}
\end{itemize}
\vspace{-8pt}
though, in all our simulation studies presented in Section \ref{sec:sims}, we evaluated the performance of NIDAAREM using the monotonicity-based switching criterion.
The monotonicity-based criterion is a natural choice as it can prevent often wasteful iterations where Nesterov increases the objective function. Both \cite{donogue2015} and \cite{giselsson2014} noted good performance from restarting Nesterov acceleration whenever this monotonicity condition is violated since an instance of non-monotonicity often indicates the momentum is leading iterates $\mathbf{x}_{k}$ in a bad direction. Moreover, as noted in \cite{donogue2015} and as we have also noted in our simulations, the first instances of non-monotonicity in Nesterov are an indication that it has entered its wave-like oscillatory phase and that the momentum is too large.

\begin{algorithm}[ht]
\caption{(NIDAAREM: DAAREM with Nesterov Initialization).}\label{euclid}
\begin{algorithmic}[1] \onehalfspacing
\State Given $\mathbf{x}_{0} \in \Omega$, a steplength $t > 0$, and integers $m \geq 1$.
\State Set $\mathbf{y}_{1} = \mathbf{x}_{0}$ and $a_{1} = 1$.
\For{k=1,2,3,...}
\State $\mathbf{x}_{k} = G_{t}( \mathbf{y}_{k} )$
\If {SWITCH$(\mathbf{x}_{k}, \mathbf{x}_{k-1}, \mathbf{y}_{k}, k) = $TRUE,} 
\State set $\tilde{\mathbf{x}} = \mathbf{x}_{k}$ and go to Step 10.
\Else 
\State $a_{k+1} = \big\{1 + \sqrt{1 + 4a_{k}^{2}} \big\}/2$
\State $\mathbf{y}_{k + 1} = \mathbf{x}_{k} + \Big( \frac{ a_{k} - 1}{a_{k+1}} \Big)(\mathbf{x}_{k} - \mathbf{x}_{k-1})$.
\EndIf
\EndFor
\State
Using $\tilde{\mathbf{x}}$ as an initial value, run DAAREM (Algorithm \ref{alg:daarem_alg}) until convergence.
\end{algorithmic}
\label{alg:nidaarem}
\end{algorithm}

The right-hand panel of Figure \ref{fig:trace_plots} shows the performance of Nesterov-initialized DAAREM (NIDAAREM) using the monotonicity-based switching criterion. This panel
shows that NIDAAREM switches from Nesterov to DAAREM after roughly 80 iterations. After the switch from Nesterov, DAAREM continues the fast initial descent of Nesterov and does not exhibit the plateau phase exhibited by DAAREM when Nesterov initialization is not used. This rapid convergence of DAAREM iterations immediately after switching from Nesterov demonstrates the potential of NIDAAREM to converge much faster than both Nesterov and DAAREM.

\vspace{-0.3cm}

\subsection{Subsetted DAAREM and NIDAAREM for Sparse Proximal Mappings}
It is common in practice to have proximal gradient mappings which are sparse -- meaning that many elements of the parameter vector are mapped directly to zero. When the proximal update is very sparse, one can potentially exploit this structure to create more efficient implementations of AA and variations of AA. Specifically, when a particular PGD scheme has sparse updates, many components of the parameter vector $\mathbf{x}_{k}$ will be repeatedly mapped to zero after several iterations. In such cases, 
the matrix $\mathbf{F}_{k}$ in the AA algorithm (Algorithm \ref{alg:restarted_aa}) 
will have many rows consisting entirely of zeros, and including these rows in the calculation of the AA update is superfluous as eliminating such rows does not change the update of the iterate $\mathbf{x}_{k}$ (see Appendix B for further details). 

Motivated by the observation that including certain rows of $\mathbf{F}_{k}$ is unnecessary, we propose and evaluate modified DAAREM and NIDAAREM procedures which only include the subset of rows of $\mathbf{F}_{k}$, where the corresponding row-sums of the absolute values of $\mathbf{F}_{k}$ are below 
a pre-specified threshold $\varepsilon_{T}$. We refer to these modified subsetting procedures as subsetted DAAREM (SDAAREM) and subsetted NIDAAREM (SNIDAAREM) respectively.
Our simulation studies in the $\ell_{1}$-penalized regression problem (see Section \ref{ss:lasso_sims}) have shown that SDAAREM and SNIDAAREM with $\varepsilon_{T} = 0.001 \times \big( p^{-1}\sum_{i=1}^{p} \sum_{j=1}^{m_{k}} |F_{ij}^{k}| \big)$, where $F_{ij}^{k}$ are the elements of $\mathbf{F}_{k}$, typically perform similarly to DAAREM/NIDAAREM in terms of the number of PG updates required for convergence.  However, in our experience, the gains in computational time from subsetting is often modest. The computational effort saved from using subsetting to reduce the dimension of the least-squares problem in Algorithm \ref{alg:restarted_aa} is often relatively small when compared with the PGD update, and the computational speed gained from solving a smaller least-squares problem is often partially offset by both the need to repeatedly compute the subset of row indices to include in each iteration.
Despite this, our experience suggests that using SDAAREM and SNIDAAREM for sparse problems such as $\ell_{1}$-penalized regression can consistently provide at least noticeable gains in speed and can be a useful tool for very sparse optimization tasks where any gains in computational time are valuable. 

\vspace{-0.5cm}

\section{Issues in Implementation}

\vspace{-0.3cm}

\subsection{Stopping Criteria}
In our numerical experiments, convergence for DAAREM and NIDAAREM is determined by the magnitude of the residual norm $|| r(\mathbf{x}_{k}) || = || G_{t}(\mathbf{x}_{k}) - \mathbf{x}_{k} ||$, and we terminate both DAAREM and NIDAAREM whenever either $|| r(\mathbf{x}_{k}) || \leq \varepsilon_{s}$ or a maximum number of iterations has been reached. We set $\varepsilon_{s} = 10^{-8}$ in each of our simulation studies. The purpose of monitoring the residual norm rather than the magnitude of change in the parameter values or the objective function is meant to avoid cases of premature termination, as early stopping could occur when DAAREM stagnates for a number of iterations. It is not unusual for DAAREM to stagnate for several iterations where there are little-to-no changes in the parameter values or objective function before converging rapidly several iterations later. Using the norm of the residuals to determine convergence prevents situations where simple stagnation could be mistaken for convergence. 
In our experience, using the stopping criterion $|| r(\mathbf{x}_{k}) || \leq 10^{-8}$ is quite strict and leads to highly accurate solutions. For lasso and penalized logistic regression, this criterion is typically stricter than other commonly used stopping criteria; for example, the criterion suggested in \cite{tibshirani2015} and used in the \verb"R" package \verb"glmnet" (\cite{Friedman2010}) of stopping when the maximum change in the absolute value of the parameters is less than $10^{-4}$. 

\vspace{-0.3cm}

\subsection{Acceleration with Warm Starts}
For many optimization problems of the form (\ref{eq:basic_form}), $h$ is indexed by a penalty parameter $\lambda$, and in such cases, one wants to compute the optimal solution $\mathbf{x}_{\lambda}^{*}$ over a range of values of $\lambda$ rather than a single choice of $\lambda$. For a given $\lambda_{1} > \ldots > \lambda_{K} \geq 0$ sequence of penalty parameters, it is often suggested that one compute the sequence of solutions $\mathbf{x}_{\lambda_{1}}^{*}, \ldots, \mathbf{x}_{\lambda_{K}}$ using ``warm starts." When using warm starts, one first computes $\mathbf{x}_{\lambda_{1}}^{*}$ and uses this as the starting value when computing $\mathbf{x}_{\lambda_{2}}^{*}$. Likewise, one uses $\mathbf{x}_{\lambda_{2}}^{*}$ as the starting value when computing $\mathbf{x}_{\lambda_{3}}^{*}$, and the remaining $\mathbf{x}_{\lambda_{4}}^{*}, \ldots, \mathbf{x}_{\lambda_{K}}^{*}$ are found by proceeding in a similar manner. 

Using warm starts with a PG scheme results in faster convergence as each starting value will be relatively close to the solution. Indeed, it has been argued that there is often little to no benefit from using Nesterov-accelerated gradient methods when using warm starts (\cite{tibshirani2015}). We further investigate the benefits of various acceleration schemes in the context of warm starts in Section 6.

\vspace{-0.5cm}

\section{Numerical Experiments} \label{sec:sims}
This section presents the results of four numerical experiments (simulation studies) 
comparing NIDAAREM with competing methods on four representative
optimization problems. The number of PGD iterations
required to converge, overall computational time, and quality
of the solution at convergence were assessed.
The performance of NIDAAREM was compared with PGD, Nesterov
acceleration, Nesterov acceleration with adaptive restarts,
DAAREM, and SQUAREM. The fixed monotonicity (fm) and 
alternating monotonicity (am) monitoring approaches
of NIDAAREM were also evaluated. 

\vspace{-0.3cm}

\subsection{$l_{1}$-penalized regression} \label{ss:lasso_sims}

Here, we present a more thorough comparison of acceleration methods for the $l_{1}$-penalized regression problem than the comparisons depicted in Figure \ref{fig:trace_plots}. For these comparisons, we generated design matrices $\mathbf{X} \in \mathbb{R}^{n \times p}$ with correlated columns by sampling the $i^{th}$ row $(X_{i1}, \ldots, X_{ip})$ of $\mathbf{X}$ from a multivariate normal distribution with mean vector $\mathbf{0}$ and $\textrm{cov}( X_{ij}, X_{ik}) = \rho^{|k-j|}$. In our simulations, we considered $\rho = 0.8$ and $\rho = 0.95$ to represent moderate and strong correlation respectively. The regression coefficients $\beta_{j}$ were set to $0$ with probability $0.8$ while the nonzero coefficients were drawn from a $t$ distribution with three degrees of freedom. We considered the following three choices of $n$ and $p$: $(n, p ) = (100, 10000), (n,p) = (10000, 100),$ and $(n, p) = (2000, 500)$ to represent scenarios where $p$ is much larger than $n$, scenarios where $n$ is much larger than $p$, and scenarios where $n$ and $p$ are roughly comparable, and we considered two values of the penalty term $\lambda$: $\lambda = 50$ and $\lambda = 500$. For each choice of $(n, p, \rho, \lambda)$, we performed $50$ simulation runs with each run having a different randomly generated $\mathbf{X}$ and $\mathbf{y}$, and for each run, we used the starting value of $\beta_{j} = 0$, for $j = 1, \ldots, p$.

Table \ref{tab:lasso_results} shows the simulation results for the small-n, large-p setting, where $(n,p) = (100, 10000)$ and $\lambda = 500$. 
Overall, NIDAAREM provides much faster convergence than the original PGD scheme and Nesterov acceleration, with NIDAAREM generating a roughly
70-fold improvement over Nesterov in terms of the median number of PGD iterations used.
Indeed, Table \ref{tab:lasso_results}
shows that, across both values of the covariate correlation $\rho$,
NIDAAREM and SNIDAAREM consistently required the fewest PGD iterations to converge. While NIDAAREM and SNIDAAREM used similar numbers of PGD iterations, SNIDAAREM had a modest, yet clear advantage in computational time, and these results demonstrate that, in sparse, high-dimensional settings, only using a subset of the parameter values in the Anderson acceleration updates can generate improvements in computational efficiency. Highlighting the ability of NIDAAREM to leverage the advantages of Nesterov with restarts and DAAREM, both versions of NIDAAREM performed substantially better than both Nesterov with restarts and DAAREM, with NIDAAREM generating 
roughly four-fold and three-fold improvements in PGD iterations over DAAREM and Nesterov with restarts respectively. For the choice of $n$ and $p$ considered in Table \ref{tab:lasso_results}, the fixed and alternating monotonicity monitoring strategies had very similar performance for both the NIDAAREM and DAAREM methods. However, there were some settings of $n$ and $p$ where the choice of monotonicity monitoring strategy led to a substantial difference in performance, and experimenting
with both strategies is recommended.
Appendix C shows additional simulation results for the other combinations of $n$, $p$, and $\lambda$.

\begin{table}[H]
\centering
\begin{tabular}{l|l rrr rrr}
\toprule
\multirow{2}{*}{$\rho$} & \multirow{2}{*}{Method} & \multicolumn{3}{c}{Number of PGD steps} &
  \multicolumn{2}{c}{Timing} &
  \multicolumn{1}{c}{$\varphi(x)$} \\
\cmidrule(r){3-5}\cmidrule(r){6-7}\cmidrule{8-8}
& & mean & median & std. dev. & mean & median & mean \\
\midrule
0.8 & PGD & 63013.7 & 59288.5 & 18825.4 & 112.5 & 105.8 & 109636.442012 \\  
  & Nesterov & 32529.7 & 31690.0 & 6805.8 & 82.5 & 80.5 & 109636.442012 \\ 
  & Nesterov (restart) & 1418.9 & 1378.0 & 267.5 & 4.6 & 4.5 & 109636.442012 \\ 
  & DAAREM (fm) & 2162.5 & 2122.5 & 315.3 & 13.4 & 13.2 & 109636.442012 \\ 
  & DAAREM (am) & 2170.7 & 2162.5 & 262.9 & 13.4 & 13.2 & 109636.442012 \\ 
  & NIDAAREM (fm) & 483.8 & 470.0 & 96.2 & 2.0 & 2.0 & 109636.442012 \\ 
  & NIDAAREM  (am) & 488.1 & \textbf{463.0} & 90.4 & 2.1 & 2.0 & 109636.442012 \\ 
  & RNIDAAREM  (am) & 513.3 & 514.5 & 97.1 & 2.2 & 1.9 & 109636.442012 \\ 
  & SDAAREM (am) & 2299.1 & 2185.0 & 654.6 & 8.2 & 7.8 & 109636.442012 \\ 
  & SNIDAAREM (am) & \textbf{482.7} & 463.5 & 101.3 & \textbf{1.6} & \textbf{1.5} & 109636.442012 \\ 
  & SQUAREM & 935.1 & 892.0 & 184.8 & 2.1 & 2.0 & 109636.442012 \\ 
\midrule
0.95 & PGD & 152756.9 & 146969.5 & 44456.0 & 272.4 & 261.2 & 115509.992958 \\ 
  & Nesterov & 62178.1 & 60538.5 & 16350.5 & 158.3 & 153.1 & 115509.992955 \\ 
  & Nesterov (restart) & 2092.1 & 2097.0 & 315.2 & 6.8 & 6.9 & 115509.992956 \\ 
  & DAAREM (fm) & 2855.6 & 2812.5 & 383.5 & 17.6 & 17.2 & 115509.992957 \\ 
  & DAAREM (am) & 2905.9 & 2962.5 & 411.9 & 17.6 & 17.2 & 115509.992957 \\ 
  & NIDAAREM (fm) & \textbf{761.1} & 699.5 & 239.5 & 3.5 & 2.9 & 115509.992957 \\ 
  & NIDAAREM (am) & 765.6 & \textbf{698.5} & 187.9 & 3.5 & 3.0 & 115509.992957 \\ 
  & RNIDAAREM (am) & 967.6 & 824.5 & 598.8 & 4.9 & 4.0 & 115509.992957 \\ 
  & SDAAREM (am) & 2990.8 & 2967.5 & 491.3 & 10.5 & 10.4 & 115509.992956 \\ 
  & SNIDAAREM (am) & 793.1 & 719.5 & 266.0 & \textbf{2.6} & \textbf{2.3} & 115509.992957 \\ 
  & SQUAREM  & 2050.6 & 1826.0 & 1277.0 & 4.6 & 4.0 & 115509.992956 \\
\bottomrule
\end{tabular}
\caption{\small{Simulation study for the $\ell_{1}$-penalized regression problem with $n = 100$, $p = 10000$, and $\lambda = 500$. Mean and median 
number of proximal gradient descent (PGD) iterations required to
converge across 50 simulation replications are shown for each method with number of iterations truncated at 500,000 if not
converged within 500,0000 iterations. Mean and median computational times across simulation replications are also shown. Mean values of the objective function at convergence are in the column labeled by $\varphi(x)$. Numbers with smallest mean/median PGD iterations and smallest mean/median times across methods are in bold text. DAAREM (fm) and DAAREM (am) denote the DAAREM method with the fixed and alternating monotonicity monitoring strategies respectively,
and RNIDAAREM denotes the NIDAAREM procedure that monitors progress in the residuals rather than the objective function. SQUAREM denotes the squared iterative acceleration method described in \cite{Varadhan2008}.} }  
\label{tab:lasso_results}
\end{table}

For the scenarios with $n = 100$ and $p=10,000$, we also compared acceleration methods in their performance on computing the full ``solution path'' when using warm starts across a decreasing sequence of penalty terms. Comparisons for computing this solution path are also presented in Appendix C.
\vspace{-0.3cm}

\subsection{Penalized logistic regression}
To evaluate the performance of NIDAAREM in accelerating convergence for $\ell_{1}$-penalized logistic regression, 
we used the {\it Madelon} data, which is a dataset available for download from the UCI machine learning
repository. This is a synthetic dataset containing 2600 observations and 500 features (i.e., $n = 2600$ and $p = 500$). 
In our numerical evaluations, we considered a sequence of $10$ penalty terms $\lambda_{1} > \lambda_{2} > \cdots > \lambda_{10}$,
where $\lambda_{1}$ is approximately equal to the smallest value for which all the associated estimated regression
coefficients are equal to zero and where the remaining penalty terms were 
set to $\lambda_{j} = \lambda_{1}(.01)^{(j-1)/9}$.

Table \ref{tab:madelon_sims} presents results for two separate values
of the penalty term, $\lambda = \lambda_{2} = 67.6$ and $\lambda = \lambda_{6} = 8.7$. Appendix C
presents results comparing performance on computing the entire solution
path across all 10 penalty terms $\lambda_{1}, \ldots, \lambda_{10}$ using warm starts. For each of the two choices of the penalty term in Table \ref{tab:madelon_sims}, each acceleration procedure was repeated 50 times where, in each replication, the initial values for the regression coefficients were drawn from a standard normal distribution.

\begin{table}[ht]
\centering
\begin{tabular}{l|l rrrrr rr}
\toprule
\multirow{2}{*}{$\lambda$} & \multirow{2}{*}{Method} & \multicolumn{3}{c}{Number of PGD steps} &
  \multicolumn{2}{c}{Timing} &
  \multicolumn{1}{c}{$\varphi(x)$} \\
\cmidrule(r){3-5}\cmidrule(r){6-7}\cmidrule{8-8}
& & mean & median & std. dev. & mean & median & mean \\
  \midrule
 67.6 & PGD & 4698.8 & 4535.5 & 589.6 & 19.08 & 19.36 & 1758.915880 \\ 
  & Nesterov & 1023.1 & 1092.5 & 341.3 & 5.70 & 6.17 & 1758.915902 \\ 
  & Nesterov (restart) & 233.7 & 234.0 & 19.0 & 1.66 & 1.66 & 1758.915880 \\ 
  & SQUAREM & 160.7 & 162.5 & 29.2 & 0.74 & 0.74 & 1758.915880 \\ 
  & DAAREM (fm) & 100.8 & 100.0 & 16.2 & 0.62 & 0.61 & 1758.915880 \\ 
  & DAAREM (am) & 100.3 & 100.0 & 14.6 & 0.63 & 0.63 & 1758.915880 \\ 
  & NIDAAREM (fm) & 73.7 & 69.5 & 19.0 & \textbf{0.50} & \textbf{0.46} & 1758.915880 \\ 
  & NIDAAREM (am) & \textbf{73.6} & \textbf{68.0} & 17.6 & 0.51 & \textbf{0.46} & 1758.915880 \\ 
  \midrule
 8.7 & PGD & 6300.6 & 6596.5 & 1070.9 & 24.86 & 26.01 & 1581.721925 \\ 
  & Nesterov & 1336.0 & 1272.5 & 493.6 & 7.32 & 6.91 & 1581.721947 \\ 
  & Nesterov (restart) & 317.9 & 323.0 & 28.8 & 2.22 & 2.25 & 1581.721925 \\ 
  & SQUAREM & 334.3 & 325.5 & 44.6 & 1.50 & 1.46 & 1581.721925 \\ 
  & DAAREM (fm) & 229.4 & 220.0 & 60.0 & 1.34 & 1.27 & 1581.721925 \\ 
  & DAAREM (am) & 193.4 & 195.0 & 16.5 & 1.16 & 1.17 & 1581.721925 \\ 
  & NIDAAREM (fm) & 189.0 & 176.5 & 64.5 & 1.18 & 1.11 & 1581.721925 \\ 
  & NIDAAREM (am) & \textbf{163.7} & \textbf{161.0} & 16.5 & \textbf{1.06} & \textbf{1.04} & 1581.721925 \\ 
\bottomrule
\end{tabular}
\caption{\small{Simulation study for the penalized logistic regression problem using the \textit{Madelon} data. Mean and median number of proximal gradient descent (PGD) iterations required to converge. Mean and median computational times across simulation replications are also shown. Mean values of the objective function at convergence are in the column labeled by $\varphi(x)$. Numbers with smallest mean/median PGD iterations and smallest mean/median times across methods are in bold text. DAAREM (fm), DAAREM (am), NIDAAREM (fm), NIDAAREM (am), and SQUAREM are as defined in Table \ref{tab:lasso_results}.}}
\label{tab:madelon_sims}
\end{table}

As shown in Table \ref{tab:madelon_sims}, 
both versions of NIDAAREM outperformed all other methods
in terms of both PGD steps performed and overall timing. The fixed and alternating monotonicity monitoring strategies of NIDAAREM had nearly identical performance when $\lambda = 67.6$, but the alternating monotonicity had modestly better performance when $\lambda$ was set to $\lambda = 8.7$. Though DAAREM performed very well relative to competing methods, 
the addition of the Nesterov initialization in NIDAAREM led to modest, but consistent improvements in convergence speed. For each choice of $\lambda$, adding restarts to Nesterov acceleration substantially improved convergence speed, with the addition of restarts leading to a roughly five-fold reduction in the number of PGD steps.
Despite the impressive gains obtained by adding restarts to Nesterov, NIDAAREM improved upon restarted Nesterov by roughly three and two-fold in the $\lambda = 67.6$ and $\lambda = 8.7$ settings respectively. Moreover, the additional timing required to perform the NIDAAREM and DAAREM steps was quite modest relative to the computational cost of performing a PGD step, and the added timing per PGD step of NIDAAREM is comparable to Nesterov acceleration with restarts.

\vspace{-0.3cm}

\subsection{Matrix Completion}
We evaluated the performance of NIDAAREM on the matrix completion problem using the \verb"movielens" 100K dataset (see e.g., \cite{harper2015}). 
This dataset contains movie ratings made by 943 individuals on 1,682 movies where each observed rating is an integer between 1 and 5. The \verb"movielens" dataset is available for download at \url{https://grouplens.org/datasets/movielens/100k/}. The \verb"movielens" dataset only contains a subset of the 1,586,126 possible ratings as each individual does not, of course, provide a rating for all the 1,682 movies listed. In fact, if one thinks of $\mathbf{A} \in \mathbb{R}^{943 \times 1682}$ as the matrix containing every possible movie rating, only $6.3\%$ percent of the entries of $\mathbf{A}$ are observed while $93.7\%$ percent of the possible movie ratings are missing.

To assess relative performance on the matrix completion problem, we first examined each method's speed in accelerating convergence when computing a sequence of ten completed matrices for the penalty terms $\lambda_{1} > \lambda_{2} > \cdots > \lambda_{10}$ using ``warm starts''.
The largest penalty term was set to $\lambda_{1} = 641$, which was chosen because it was the smallest integer penalty that yielded a solution consisting entirely of zeros. The smallest penalty term $\lambda_{10}$ was set to $\lambda_{10} = 6.41$, and the remaining values of $\lambda_{k}$ were set to be equally spaced on the log scale, i.e., $\log(\lambda_{k}) = \log(\lambda_{k-1}) - \log(100)/9$, for $k = 2, \ldots, 10$.

The results from the matrix completion problem with warm starts are shown in Table \ref{tab:matrixcomplete_warm},
where the total number of PGD steps and total time needed to 
compute the entire solution path of 10 matrices is shown.
As shown in Table \ref{tab:matrixcomplete_warm}, this is an example where applying Nesterov's momentum method with warm starts does not work well, and in fact, Nesterov's method leads to moderately slower convergence than using PGD itself. 
Adding restarts to Nesterov substantially improved the number of PGD steps needed to compute the full solution path, but the reduction in computation time was not as dramatic, because Nesterov with restarts requires an objective function evaluation in every iteration. Both DAAREM and NIDAAREM performed very well on this problem,
with DAAREM delivering a roughly four-fold improvement over restarted Nesterov in the number of PGD steps required for convergence. Both versions of NIDAAREM considered (both fixed monotonicity monitoring and residual-based alternating monotonicity monitoring) required the same number of PGD steps, with NIDAAREM requiring roughly 20 percent fewer PGD iterations than that of DAAREM.
Overall, the results in Table \ref{tab:matrixcomplete_warm} indicate that, while using unaccelerated PGD together with warm starts can often be an efficient approach that Nesterov does not improve upon, the use of effective acceleration schemes can still generate notable improvements over PGD in computational speed.

\begin{table}[ht]
\centering
\begin{tabular}{lrr}
\toprule
Method & Number of PGD steps  & Time (in seconds) \\ 
\midrule
PGD & 14593 & 56970.7 \\ 
  Nesterov & 18278 & 124236.6 \\ 
  Nesterov w restarts & 3485 & 33970.1 \\ 
  SQUAREM & 2565 & 14192.1 \\ 
  DAAREM (fm) & 880 & \textbf{6425.8} \\ 
  NIDAAREM (am) & \textbf{731} & 8394.3 \\ 
  RNIDAAREM (am) & \textbf{731} & 8370.2 \\ 
  Soft-Impute & NA  & 184234.5 \\ 
   \bottomrule
\end{tabular}
\caption{{\small Results from performing matrix completion on the \textbf{movielens} data with warm starts over a sequence of $10$ penalty terms. The total number of PGD steps required to compute the entire sequence of matrix completion solutions is shown for each method. In addition, the time
required to compute the entire sequence of matrix completion solutions is shown. DAAREM (fm), NIDAAREM (am), and SQUAREM are as defined in Table \ref{tab:lasso_results}.
Soft-Impute refers to the iterative algorithm described in \cite{Mazumder2010}, and RNIDAAREM (am) denotes the NIDAAREM procedure with alternating monotonicity that monitors progress in the residuals rather than the objective function.}} 
\label{tab:matrixcomplete_warm}
\end{table}

\begin{table}[ht]
\centering
\begin{tabular}{l|l rrr rrr}
\toprule
\multirow{2}{*}{$\lambda$} & \multirow{2}{*}{Method} & \multicolumn{3}{c}{Number of PGD steps} &
  \multicolumn{2}{c}{Timing} &
  \multicolumn{1}{c}{$\varphi(x)$} \\
\cmidrule(r){3-5}\cmidrule(r){6-7}\cmidrule{8-8}
& & mean & median & std. dev. & mean & median & mean \\
\midrule
20 & PGD & 2493.0 & 2493.0 & 0.0 & 9755.1 & 9688.7 & -107853.042114 \\ 
   & Nesterov & 3304.0 & 3304.0 & 0.0 & 22606.3 & 22291.4 & -107853.042114 \\ 
   & Nesterov w restarts & 545.0 & 541.0 & 24.7 & 5265.7 & 5174.1 & -107853.042114 \\ 
   & SQUAREM & 951.4 & 948.0 & 21.4 & 5366.3 & 5337.6 & -107853.042114 \\ 
   & DAAREM (am) & 165.0 & 165.0 & 2.4 & \textbf{1218.3} & \textbf{1205.0} & -107853.042114 \\ 
   & NIDAAREM (am) & \textbf{122.0} & \textbf{122.5} & 3.5 & 1325.0 & 1337.0 & -107853.042114 \\ 
   & RNIDAAREM (am) & \textbf{122.0} & \textbf{122.5} & 3.5 & 1322.6 & 1326.8 & -107853.042114 \\ 
   & SoftImpute &  NA & NA & NA & 7520.9 & 7449.3 & -107853.042674 \\ 
\bottomrule
\end{tabular}
\caption{{\small Results from the matrix completion simulation study based on the \textbf{movielens} data
with a fixed penalty term set to $\lambda = 20$. This table shows summary performance results across $10$ different random starting values. DAAREM (am), NIDAAREM (am), and SQUAREM are as defined in Table \ref{tab:lasso_results}. Soft-Impute refers to the iterative algorithm described in \cite{Mazumder2010}, and RNIDAAREM (am) denotes the NIDAAREM procedure with alternating monotonicity that monitors progress in the residuals rather than the objective function.}} 
\label{tab:movielens_results}
\end{table}

Table \ref{tab:movielens_results} displays convergence speed summaries for the movielens matrix completion problem with a fixed penalty term of $\lambda = 20$ across 10 different
random starting values, and Appendix C presents 
results on the matrix completion problem for $\lambda = 200$. For each of the 10 runs, starting values were
drawn from a Gaussian distribution with mean and standard deviation set to the sample mean and standard deviation of the observed movie ratings respectively.
For a fixed value of $\lambda = 20$, the relative performance of the acceleration methods considered
closely matches that of matrix completion with warm starts. This indicates that using starting values relatively close to the solution, as is done in the warm starts example, does not greatly diminish the benefits of using acceleration schemes such as NIDAAREM or Nesterov with restarts -- at least for the convergence criterion that we have used.

\vspace{-0.3cm}

\subsection{Quadratic Programming with Constraints}
For these simulations, we study the following minimization problem: 
\begin{eqnarray}
&&\textrm{minimize } \frac{1}{2}\mathbf{x}^{T}\mathbf{Q}\mathbf{x} + \mathbf{q}^{T}\mathbf{x} \qquad \textrm{subject to } -1 \leq \mathbf{x} \leq 1,  \nonumber      
\end{eqnarray}
where $\mathbf{Q} \in \mathbb{R}^{p \times p}$ is a positive definite matrix, $\mathbf{q} \in \mathbb{R}^{p}$, and $\mathbf{x} \in \mathbb{R}^{p}$.
The constraint $-1 \leq \mathbf{x} \leq 1$ means that $-1 \leq x_{j} \leq 1$ for each element $x_{j}$ of $\mathbf{x}$. This minimization problem can be expressed in the form (\ref{eq:basic_form}) by defining $g(\mathbf{x}) = \frac{1}{2}\mathbf{x}^{T}\mathbf{Q}\mathbf{x} + \mathbf{q}^{T}\mathbf{x}$ and $h(\mathbf{x})$ as $h(\mathbf{x}) = 0$ if $x \in \mathcal{C}$ and $h(\mathbf{x}) = \infty$ if $\mathbf{x} \not\in \mathcal{C}$, where $\mathcal{C} = \{\mathbf{x} \in \mathbb{R}^{p}: -1 \leq \mathbf{x} \leq 1\}$. Recalling (\ref{eq:prox_mapping}), the PG update with steplength $t$ for this problem is
\begin{eqnarray}
\mathbf{x}_{k+1} &=& \argmin_{\mathbf{z} \in \mathbb{R}^{p}} \Big\{ \frac{1}{2}|| \mathbf{z} - \mathbf{x}_{k} + t\mathbf{Q}\mathbf{x}_{k} + t\mathbf{q}  ||^{2} + t h(\mathbf{z}) \Big\}  \nonumber \\
&=& \mathbf{1}\{ \mathbf{b}_{k}^{t} > 1 \} - \mathbf{1}\{ \mathbf{b}_{k}^{t} < -1 \} + \mathbf{b}_{k}^{t} \circ \mathbf{1}\{ \mathbf{b}_{k}^{t} \in \mathcal{C} \}, \nonumber
\end{eqnarray}
where $\mathbf{b}_{k}^{t} = \mathbf{x}_{k} - t\mathbf{Q}\mathbf{x}_{k} - t\mathbf{q}$ and $\circ$ denotes the Hadamard product. The term $\mathbf{1}\{ \mathbf{b}_{k}^{t} \in \mathcal{C} \}$ 
denotes the $p \times 1$ vector whose elements equal one if the corresponding element of $\mathbf{b}_{k}^{t}$ belongs to $\mathcal{C}$ and equal zero otherwise, and the vectors $\mathbf{1}\{ \mathbf{b}_{k}^{t} > 1 \}$ and $\mathbf{1}\{ \mathbf{b}_{k}^{t} < -1 \}$ are defined similarly.

We generated $\mathbf{Q}$ as $\mathbf{Q} = \mathbf{U}\mathbf{D}\mathbf{U}^{T}$, where $\mathbf{U} = (\mathbf{B}^{T}\mathbf{B})^{-1/2}$ and 
$\mathbf{B} \in \mathbb{R}^{p \times p}$ was generated by drawing the elements of $\mathbf{B}$ from a 
Gaussian distribution with mean $0$ and standard deviation $1$. 
To ensure the condition number of $\mathbf{Q}$ was equal to $\kappa$, we defined the matrix 
$\mathbf{D}$ as $\mathbf{D} = \textrm{diag}\{ d_{1}, \ldots, d_{p} \}$
where $\log d_{j} = \log(\kappa)\{1 - (j-1)/(p-1)\}$. 
We generated the elements $q_{j}$ of the vector $\mathbf{q}$ as 
$q_{j} = \varepsilon_{j} \{ \tfrac{1}{p-1} \sum_{i=1}^{p}(Q_{ij} - \bar{Q}_{.j})^{2}\}^{1/2}$,
where $\bar{Q}_{.j} = \tfrac{1}{p}\sum_{i=1}^{p} Q_{ij}$ and $\varepsilon_{j}$ are independent, mean zero Gaussian random variables with standard deviation $1/5$.


\begin{table}[ht]
\centering
\begin{tabular}{l rrr rrr}
\toprule
 \multirow{2}{*}{Method} & \multicolumn{3}{c}{Number of PGD steps} &
  \multicolumn{2}{c}{Timing} &
  \multicolumn{1}{c}{$\varphi(x)$} \\
\cmidrule(r){2-4}\cmidrule(r){5-6}\cmidrule{7-7}
 & mean & median & std. dev. & mean & median & mean \\
\midrule
PGD & 117638.1 & 117675.0 & 11805.5 & 118.4 & 118.2 & -5851571.355630 \\ 
  Nesterov & 71069.0 & 70270.0 & 6177.3 & 143.0 & 141.0 & \textbf{-5851571.356494} \\ 
  Nesterov w restarts & 125503.2 & 124989.0 & 12590.9 & 374.3 & 373.1 & -5851571.356143 \\ 
  SQUAREM & 35769.8 & 35679.0 & 2732.1 & 59.5 & 59.5 & -5851571.356000 \\ 
  DAAREM (am) & 6571.1 & 6505.0 & 1024.3 & 17.6 & 17.7 & -5851571.355633 \\ 
  NIDAAREM (fm) & \textbf{2937.9} & \textbf{2920.0} & 208.1 & \textbf{7.1} & \textbf{7.0} & -5851571.356041 \\ 
  NIDAAREM (am) & 6659.4 & 6645.0 & 932.4 & 17.9 & 17.7 & -5851571.355634 \\   
\bottomrule
\end{tabular}
\caption{{\small Simulation results for the constrained quadratic programming simulation study. Summary performance measures across $100$ different randomly generated $\mathbf{Q}$ and $\mathbf{q}$ are shown. DAAREM (am), NIDAAREM (fm), NIDAAREM (am), and SQUAREM are as defined in Table \ref{tab:lasso_results}}} 
\label{tab:qp_results}
\end{table}

Performance measures of different acceleration methods across 100 randomly generated matrices $\mathbf{Q}$ and vectors $\mathbf{q}$ in the constrained quadratic programming
problem are reported in Table \ref{tab:qp_results}. As shown in this table, NIDAAREM with fixed
monotonicity control consistently required fewer PGD steps to converge than 
competing methods. While both DAAREM and NIDAAREM with alternating monotonicity control provided a ten-fold improvement
over Nesterov acceleration in terms of PG steps, 
this is an example where the fixed monotonicity monitoring strategy of DAAREM is
the clearly superior approach to tracking progress in the objective function. 
It is also interesting to note that this is the only example where we have observed that Nesterov
with restarts performs worse than Nesterov without restarts, possibly driven, in part, 
by the discontinuities in the PG updates. There were small differences
in the mean value of the objective function at convergence, with Nesterov achieving the
best average objective across simulation replications. These differences were quite small, and
NIDAAREM with fixed monotonicity monitoring had an average objective function value 
very close to that of Nesterov and superior to both PGD
and NIDAAREM with alternating monotonicity monitoring.

\vspace{-0.5cm}

\section{Conclusion}
\label{sec:conc}
In this paper, we have utilized and further developed a version of Anderson acceleration and combined it with Nesterov's method in a two-phase acceleration procedure. This hybrid procedure was constructed to take advantage of the strengths of each acceleration scheme in the particular context of proximal gradient algorithms. 
DAAREM and other variations of Anderson acceleration have shown impressive performance in the context of accelerating EM or MM algorithms and other fixed point iterations, and they also provide fast local convergence for many proximal gradient algorithms. However, even when DAAREM delivers substantial acceleration of a proximal gradient scheme, it often exhibits a long ``plateau phase'' before suddenly converging quickly. In contrast, Nesterov acceleration often delivers impressive reductions early on while subsequently exhibiting oscillatory behavior and slower convergence. Starting with a Nesterov initialization phase allows one to often bypass the earlier slow phase of DAAREM while taking advantage of the more rapid local convergence provided by DAAREM (this behavior is well illustrated in Figure \ref{fig:trace_plots}). As such, the two-phase strategy of NIDAAREM delivers impressive speed-ups of proximal gradient algorithms. 

Our strategy for choosing the monotonicity tolerance term $\varepsilon_k$ is admittedly somewhat ad-hoc, even though it has performed well in all our experiments and is closely related to the monotonicity monitoring strategy used in DAAREM.  Nevertheless, future work should explore more principled approaches for specifying $\varepsilon_{k}$. One approach, for example, would be to allow $\varepsilon_k$ to be adaptive based on the distance to the minimum, i.e. $\varepsilon_k = c \,(\varphi(\mathbf{x}_{k-1}) - \varphi(\mathbf{x}^*)),$ where $0 \le c < 1$ is a pre-specified scaling constant
and $\varphi(\mathbf{x}^{*})$ is the minimum value of the objective function. Because $\varphi(\mathbf{x}^*)$ is unknown, we could use Aitken's extrapolation of previous objective function values to generate an estimate of $\varphi(\mathbf{x}^*)$.

All the acceleration schemes and numerical experiments presented here assume the proximal gradient method uses a fixed step length that does not vary across iterations. However, because finding a good fixed step length that guarantees descent can be a challenge in many applications, it would be worthwhile to more systematically explore the use of NIDAAREM when the step size is chosen adaptively every iteration. One reason for only considering a fixed step length is that Anderson acceleration is designed for extrapolation when there is a single fixed point function under consideration. We have also performed a few preliminary experiments where the step size is varied and is chosen to be larger than the step size that guarantees descent of PGD. In these experiments, we have found that NIDAAREM converges across a broad range of steplengths but that larger steplengths often make a modest, but notable improvement in the convergence speed of NIDAAREM. These results imply that the gains in convergence speed come mostly from the NIDAAREM acceleration scheme and that the larger step sizes play a lesser role, and hence, because the extent to which one can increase 
the steplength and maintain the robustness of NIDAAREM is substantially problem-dependent,
we recommend one use a steplength which guarantees descent unless one has strong evidence
that a larger steplength has robust performance with NIDAAREM.

We developed and evaluated NIDAAREM in accelerating proximal gradient descent schemes due to the frequently observed fast descent of Nesterov acceleration in this context. However, NIDAAREM could be easily extended to accelerate EM and MM algorithms or more general fixed point iterations. Adapting Nesterov's momentum method for use in accelerating fixed point iterations
is straightforward (\cite{tran2024}), and Nesterov acceleration has demonstrated effectiveness in accelerating procedures not based on proximal gradient descent (e.g., \cite{ye2020, tang2023}). Because DAAREM can be directly applied to general fixed point iterations, one could directly use a modified Nesterov acceleration in the initial phase followed by DAAREM after Nesterov has violated a monotonicity condition. Evaluating such a modified NIDAAREM procedure on more general types of fixed-point iterations would be an interesting topic to examine in future research.

\medskip
\begin{center}
{\large\bf Supplementary Material}
\end{center}
\vspace{-.6cm}

\begin{description}
\item[R-package:] An R-package \verb"nidaarem" performing the methods described in the article is available at \url{https://github.com/nchenderson/nidaarem}. 

\item[Simulation code:] All \textbf{R} code and saved simulation results needed to 
reproduce the simulation results shown in Figure 1, Tables 1-5,
and simulation results shown in the appendices can be found at \url{https://github.com/nchenderson/nidaarem_reproduce}
\end{description}
\vspace{-.6cm}

\bibliographystyle{agsm}
\bibliography{sdaarem}

@article{agarwal2024,
  title={Quasi-Newton Acceleration of {EM} and {MM} Algorithms via {B}royden’s Method},
  author={Agarwal, Medha and Xu, Jason},
  journal={Journal of Computational and Graphical Statistics},
  volume={33},
  number={2},
  pages={393--406},
  year={2024}
}

@article{Anderson1965,
 author = {Anderson, Donald G.},
 title = {Iterative Procedures for Nonlinear Integral Equations},
 journal = {J. ACM},
 issue_date = {Oct. 1965},
 volume = {12},
 number = {4},
 year = {1965},
 pages = {547--560}
}

@article{Beck2009,
  title={A fast iterative shrinkage-thresholding algorithm for linear inverse problems},
  author={Beck, Amir and Teboulle, Marc},
  journal={SIAM journal on imaging sciences},
  volume={2},
  number={1},
  pages={183--202},
  year={2009},
  publisher={SIAM}
}

@book{Beck2017,
  title={First-order methods in optimization},
  author={Beck, Amir},
  volume={25},
  year={2017},
  publisher={SIAM}
}

@article{Brezinski2019,
  title={The genesis and early developments of {A}itken’s process, {S}hanks’ transformation, the $\varepsilon$--algorithm, and related fixed point methods},
  author={Brezinski, Claude and Redivo--Zaglia, Michela},
  journal={Numerical Algorithms},
  volume={80},
  number={1},
  pages={11--133},
  year={2019}
}

@Article{Candes2009,
author={Cand{\`e}s, Emmanuel J. and Recht, Benjamin},
title={Exact Matrix Completion via Convex Optimization},
journal={Foundations of Computational Mathematics},
year={2009},
volume={9},
number={6},
pages={717-772}
}

@article{donogue2015,
  title={Adaptive restart for accelerated gradient schemes},
  author={O{D}onoghue, Brendan and Candes, Emmanuel},
  journal={Foundations of computational mathematics},
  volume={15},
  number={3},
  pages={715--732},
  year={2015},
  publisher={Springer}
}

@article{Fang2009,
	Author = {Hawren Fang and Yousef Saad},
	Journal = {Numerical Linear Algebra with Applications},
	Pages = {197--221},
	Title = {Two classes of multisecant methods for nonlinear acceleration},
	Volume = {16},
    Number = {3},
	Year = {2009}
}

@article{Friedman2010,
  title={Regularization paths for generalized linear models via coordinate descent},
  author={Friedman, Jerome and Hastie, Trevor and Tibshirani, Rob},
  journal={Journal of statistical software},
  volume={33},
  number={1},
  pages={1-22},
  year={2010}
}

@article{harper2015,
  title={The movielens datasets: History and context},
  author={Harper, F Maxwell and Konstan, Joseph A},
  journal={{ACM} transactions on interactive intelligent systems (tiis)},
  volume={5},
  number={4},
  pages={1--19},
  year={2015}
}

@inproceedings{giselsson2014,
  title={Monotonicity and restart in fast gradient methods},
  author={Giselsson, Pontus and Boyd, Stephen},
  booktitle={53rd IEEE Conference on Decision and Control},
  pages={5058--5063},
  year={2014},
  organization={IEEE}
}

@article{henderson2019,
  title={Damped {A}nderson acceleration with restarts and monotonicity control for accelerating {EM} and {EM}-like algorithms},
  author={Henderson, Nicholas C and Varadhan, Ravi},
  journal={Journal of Computational and Graphical Statistics},
  volume = {28},
  number = {4},
  pages={834--846},
  year={2019}
}

@book{Kelley1999,
  title={Iterative methods for optimization},
  author={Kelley, Carl T},
  year={1999},
  publisher={{SIAM}}
}

@inproceedings{kinga2015,
  title={Adam: {A} method for stochastic optimization},
  author={Kinga, Diederik P and Ba, Jimmy},
  booktitle={International conference on learning representations (ICLR)},
  volume={5},
  number={6},
  year={2015}
}

@article{kuroda2023,
  title={Acceleration of the {EM} algorithm},
  author={Kuroda, Masahiro and Geng, Zhi},
  journal={Wiley Interdisciplinary Reviews: Computational Statistics},
  volume={15},
  number={6},
  pages={e1618},
  year={2023}
}

@article{Levy2018,
  title={Online adaptive methods, universality and acceleration},
  author={Levy, Kfir Y and Yurtsever, Alp and Cevher, Volkan},
  journal={Advances in neural information processing systems},
  volume={31},
  year={2018}
}

@article{lin2018,
  title={Catalyst acceleration for first-order convex optimization: from theory to practice},
  author={Lin, Hongzhou and Mairal, Julien and Harchaoui, Zaid},
  journal={Journal of Machine Learning Research},
  volume={18},
  number={212},
  pages={1--54},
  year={2018}
}

@article{Louis1982,
  title={Finding the observed information matrix when using the {EM} algorithm},
  author={Louis, Thomas A},
  journal={Journal of the Royal Statistical Society Series B: Statistical Methodology},
  volume={44},
  number={2},
  pages={226--233},
  year={1982}
}

@article{Mazumder2010,
  title={Spectral regularization algorithms for learning large incomplete matrices},
  author={Rahul Mazumder and Trevor Hastie and Robert Tibshirani},
  journal={Journal of machine learning research},
  volume={11},
  number={Aug},
  pages={2287--2322},
  year={2010}
}

@inproceedings{Nesterov1983,
  title={A method for solving the convex programming problem with convergence rate O (1/k\^{} 2)},
  author={Nesterov, Yurii E},
  booktitle={Dokl. akad. nauk Sssr},
  volume={269},
  pages={543--547},
  year={1983}
}

@book{nocedal2006,
  title={Numerical optimization},
  author={Nocedal, Jorge and Wright, Stephen},
  year={2006},
  publisher={Springer Science \& Business Media}
}

@article{parikh2014,
  title={Proximal Algorithms},
  author={Parikh, Neal and Boyd, Stephen},
  journal={Foundations and Trends in Optimization},
  volume={1},
  number={3},
  pages={127--239},
  year={2014}
}

@article{Pratapa2015,
 author = {Phanisri R. Pratapa and Phanish Suryanarayana},
 journal = {Chemical Physical Letters},
 pages = {69-74},
 title = {Restarted {P}ulay mixing for efficient and robust acceleration of fixed-point iterations},
 volume = {635},
 year = {2015}
}

@inproceedings{roulet2017,
  title={Sharpness, restart and acceleration},
  author={Roulet, Vincent and d'Aspremont, Alexandre},
  booktitle={Advances in Neural Information Processing Systems},
  pages={1119--1129},
  year={2017}
}

@article{Saad1986,
author = {Youcef Saad and Martin H. Schultz},
title = {{GMRES:} A Generalized Minimal Residual Algorithm for Solving Nonsymmetric Linear Systems},
journal = {SIAM Journal on Scientific and Statistical Computing},
volume = {7},
number = {3},
pages = {856-869},
year = {1986}
}

@inproceedings{scieur2024,
  title={Adaptive {Q}uasi-{N}ewton and {A}nderson acceleration framework with explicit global (accelerated) convergence rates},
  author={Scieur, Damien},
  booktitle={International Conference on Artificial Intelligence and Statistics},
  pages={883--891},
  year={2024}
}

@article{Smith1987,
 author = {David A. Smith and William F. Ford and Avram Sidi},
 journal = {SIAM Review},
 pages = {199-233},
 title = {Extrapolation Methods for Vector Sequences},
 volume = {29},
 issue = {2},
 year = {1987}
}

@article{su2016,
  title={A differential equation for modeling {N}esterov's accelerated gradient method: {T}heory and insights},
  author={Su, Weijie and Boyd, Stephen and Candes, Emmanuel J},
  journal={Journal of Machine Learning Research},
  volume={17},
  number={153},
  pages={1--43},
  year={2016}
}

@inproceedings{sutskever2013,
  title={On the importance of initialization and momentum in deep learning},
  author={Sutskever, Ilya and Martens, James and Dahl, George and Hinton, Geoffrey},
  booktitle={International conference on machine learning},
  pages={1139--1147},
  year={2013},
  organization={pmlr}
}

@article{tang2023,
  title={Accelerating Fixed-Point Algorithms in Statistics and Data Science: A State-of-Art Review},
  author={Tang, Bohao and Henderson, Nicholas C and Varadhan, Ravi},
  journal={Journal of Data Science},
  volume={21},
  number={1},
  year={2023}
}

@article{themelis2019,
  title={Super{M}ann: a superlinearly convergent algorithm for finding fixed points of nonexpansive operators},
  author={Themelis, Andreas and Patrinos, Panagiotis},
  journal={IEEE Transactions on Automatic Control},
  volume={64},
  number={12},
  pages={4875--4890},
  year={2019}
}

@book{tibshirani2015,
  title={Statistical learning with sparsity: the lasso and generalizations},
  author={Tibshirani, Robert and Wainwright, Martin and Hastie, Trevor},
  year={2015}
}

@article{toth2015,
  title={Convergence analysis for {A}nderson acceleration},
  author={Toth, Alex and Kelley, Carl T},
  journal={SIAM Journal on Numerical Analysis},
  volume={53},
  number={2},
  pages={805--819},
  year={2015},
  publisher={SIAM}
}

@article{tran2024,
  title={From {H}alpern’s fixed-point iterations to {N}esterov’s accelerated interpretations for root-finding problems},
  author={Tran-Dinh, Quoc},
  journal={Computational Optimization and Applications},
  volume={87},
  number={1},
  pages={181--218},
  year={2024}
}

@article{Varadhan2008,
	Author = {Ravi Varadhan and Christophe Roland},
	Journal = {Scandinavian Journal of Statistics},
	Pages = {335--353},
	Title = {Simple and Globally convergent methods for accelerating the convergence of any {EM} algorithm},
	Volume = {35},
    Number = {2},
	Year = {2008}
}

@article{Walker2011,
	Author = {Homer F. Walker and Peng Ni},
	Journal = {SIAM Journal on Numerical Analysis},
	Pages = {1715--1735},
	Title = {Anderson acceleration for fixed-point iterations},
	Volume = {49},
    Number = {4},
	Year = {2011}
 }

@article{ye2020,
  title={Nesterov's acceleration for approximate {N}ewton},
  author={Ye, Haishan and Luo, Luo and Zhang, Zhihua},
  journal={Journal of Machine Learning Research},
  volume={21},
  number={142},
  pages={1--37},
  year={2020}
}

@article{zhang2020,
  title={Globally convergent type-{I} {A}nderson acceleration for nonsmooth fixed-point iterations},
  author={Zhang, Junzi and O'Donoghue, Brendan and Boyd, Stephen},
  journal={SIAM Journal on Optimization},
  volume={30},
  number={4},
  pages={3170--3197},
  year={2020}
}

\newpage 

\appendix
\setcounter{table}{0}
\renewcommand{\thetable}{S\arabic{table}}

\section{Residual-based Acceptance Criterion}
In the $k^{th}$ iteration of the DAAREM algorithm, the residual of the iterate $\mathbf{x}_{k}$ is defined as $r(\mathbf{x}_{k}) = G_{t}( \mathbf{x}_{k} ) - \mathbf{x}_{k}$. 
Using $(\rho_{k}, \gamma)$ rather than $\varepsilon_{k}$ (as is used in monotonicity control) to denote the control parameters in this context, the residual-based criterion that we employ is the following
\begin{equation}
\textrm{Accept}( \mathbf{y}_{k+1}, \mathbf{x}_{k}, \rho_{k}, \gamma)    = \begin{cases}
\textrm{TRUE} & \textrm{if } || r(\mathbf{y}_{k+1}) || \leq \min\{ || r(\mathbf{x}_{k}) || + || r( \mathbf{x}_{0} )||\rho_{k}^{k},
K|| r(\mathbf{x}_{0}) || g_{\gamma}(k) \} \\
\textrm{FALSE} & \textrm{otherwise, }
\end{cases}
\label{eq:mon_cond_residual}
\end{equation}
where, in (\ref{eq:mon_cond_residual}), $r( \mathbf{x}_{0} ) = G_{t}( \mathbf{x}_{0} ) - \mathbf{x}_{0}$ and $\rho_{k} \in [0, 1)$. 
As with the alternating monotonicity approach for varying $\varepsilon_{k}$, we vary $\rho_{k}$ between $0$ and a fixed, positive number $\rho \in (0,1)$
across cycles.

Condition (\ref{eq:mon_cond_residual}) requires that at least $|| r(\mathbf{y}_{k+1})||/ || r(\mathbf{x}_{k}) || \leq 1 + \rho_{k}^{k} || r( \mathbf{x}_{0} ) ||/ || r(\mathbf{x}_{k})||$ is satisfied. In other words, the ratio between the residual norm associated with the proposed extrapolated $\mathbf{y}_{k+1}$ and the residual norm associated with $\mathbf{x}_{k}$ is allowed to be greater than $1$, but the amount by which $|| r(\mathbf{y}_{k+1})||/|| r(\mathbf{x}_{k}) ||$ can exceed $1$ is limited by the magnitude of $\rho_{k}^{k}$ times the ratio between the initial and the current residual norm. For the early cycles which have $\rho_{k} = \rho$, the restriction on the residual norm ratio $|| r(\mathbf{y}_{k+1}) ||/|| r( \mathbf{x}_{k} ) ||$ will be loose as $\rho_{k}^{k}|| r(\mathbf{x}_{0}) ||/||r(\mathbf{x}_{k})||$ will be relatively large, but the restriction on this ratio will be much closer to $1$ closer to convergence. In all our simulation studies, our default choice is $\rho = 0.95$.
Condition (\ref{eq:mon_cond_residual}) also requires the norm of the residual for the extrapolated iterate $\mathbf{y}_{k+ 1}$
be less than $K || r(\mathbf{x}_{0}) || g_{\gamma}(k)$. Here, $g_{\gamma}(k)$ is defined as $g_{\gamma}(k) = (1 + n_{aa})^{-(1 + \gamma)}$, where $n_{aa}$ is the number of Anderson extrapolations accepted up to iteration $k$ and $\gamma > 0$. Using $g_{\gamma}(k)$ as an additional requirement adds an extra layer of robustness to the algorithm as it is very similar to the acceptance criterion described in \cite{zhang2020} in the context of ensuring global convergence of a type-I Anderson acceleration scheme. Our residual-based acceptance criterion is also similar to the acceptance criterion suggested by 
\cite{themelis2019} in the context of a proposed robust Newton-type procedure for finding the fixed point of a function.

Line 14 of the DAAREM algorithm in the main manuscript also includes an extra check that monitors progress after each ``cycle'' is completed, namely, a series of $m$ steps starting from order-1 through order-m AA extrapolation. Though adding this cycle-level check does not change performance in most cases, including this cycle-level check can improve robustness in some cases as this adds extra damping whenever sufficient progress is not observed over the cycle. In our implementation, we use a direct monotonicity criterion; namely, the cycle accept criterion is true if $\varphi(\mathbf{x}_{k+1}) \leq \varphi^{*}$. Hence, extra damping is applied whenever the objective function does not improve over a single cycle.

\section{Sparse Proximal Mappings and Subsetted DAAREM updates}
As described in Section 4.2 of the main manuscript, when running a sparse proximal gradient scheme or an accelerated proximal gradient scheme many components of $\mathbf{x}_{k}$ will typically be repeatedly mapped to zero after several iterations. Thus, after several iterations of Anderson acceleration,  $\mathbf{X}_{k}$ and $\mathbf{F}_{k}$ will often have many rows which consist entirely of zeros. Including such rows in the computation of the DAAREM update is redundant as eliminating such rows will lead to the same update for $\mathbf{x}_{k}$. Though a direct exercise to prove, this is stated formally in the following proposition.

\noindent
{\bf Proposition S1.} {\it Let $\mathbf{F}_{k} \in \mathbb{R}^{p \times m_{k}}$ and $\mathbf{f}_{k} \in \mathbb{R}^{p}$ be as defined in Algorithm 2 of the main manuscript. Let $F_{ij}^{k}$ be the $(i,j)$ element of $\mathbf{F}_{k}$
and let $\mathcal{I}_{k}$ be the set of indices where the rows of $\mathbf{F}_{k}$ have at least one non-zero entry, i.e., $\mathcal{I}_{k} = \big\{ i \in \{1, \ldots, p\}: \sum_{j=1}^{m_{k}} |F_{ij}^{k}| > 0 \big\}$. Also, let $\mathbf{F}_{k,\mathcal{I}_{k}}$ denote the sub-matrix of $\mathbf{F}_{k}$ that only includes the rows of $\mathbf{F}_{k}$ that are in $\mathcal{I}_{k}$ and let $\mathbf{f}_{k,\mathcal{I}_{k}}$ denote the vector that only includes the elements of $\mathbf{f}_{k}$ that are in $\mathcal{I}_{k}$. Then,
\begin{equation}
\mathbf{F}_{k}^{T}\mathbf{F}_{k} = \mathbf{F}_{k,\mathcal{I}_{k}}^{T}\mathbf{F}_{k,\mathcal{I}_{k}} \quad \textrm{ and }  \quad \mathbf{F}_{k}^{T}\mathbf{f}_{k} = \mathbf{F}_{k,\mathcal{I}_{k}}^{T}\mathbf{f}_{k, \mathcal{I}_{k}} \nonumber 
\end{equation}
and the Anderson acceleration of $\mathbf{x}_{k}$ is given by
\begin{eqnarray}
\mathbf{x}_{k+1} &=& \mathbf{x}_{k} + \mathbf{f}_{k} - (\mathbf{X}_{k} + \mathbf{F}_{k})\bgamma^{(k)} \nonumber \\
&=& \mathbf{x}_{k} + \mathbf{f}_{k} - (\mathbf{X}_{k} + \mathbf{F}_{k})(\mathbf{F}_{k,\mathcal{I}_{k}}^{T}\mathbf{F}_{k,\mathcal{I}_{k}})^{-1}\mathbf{F}_{k,\mathcal{I}_{k}}^{T}\mathbf{f}_{k, \mathcal{I}_{k}}. \nonumber
\end{eqnarray}
}
\noindent
\textit{Proof of Proposition S1:}
The $(i,j)$ component of the $m_{k} \times m_{k}$ matrix $\mathbf{F}_{k}^{T}\mathbf{F}_{k}$ is
\begin{eqnarray}
[\mathbf{F}_{k}^{T}\mathbf{F}_{k}]_{i,j} &=& \sum_{h=1}^{p} F_{hi}^{k} F_{hj}^{k} \nonumber \\
&=& \sum_{h=1}^{p} I(h \in \mathcal{I}_{k}) F_{hi}^{k} F_{hj}^{k}
+ \sum_{h=1}^{p} I(h \in \mathcal{I}_{k}^{c}) F_{hi}^{k} F_{hj}^{k} \nonumber \\
&=& \sum_{h=1}^{p} I(h \in \mathcal{I}_{k}) F_{hi}^{k} F_{hj}^{k}. \nonumber 
\end{eqnarray}
where the last equality follows from the fact that $F_{hi}^{k} = F_{hj}^{k} = 0$ whenever
$h \in \mathcal{I}_{k}^{c}$. The $(i,j)$ component of the 
$m_{k} \times m_{k}$ matrix $\mathbf{F}_{k, \mathcal{I}_{k}}^{T}\mathbf{F}_{k, \mathcal{I}_{k}}$ is
\begin{eqnarray}
[\mathbf{F}_{k, \mathcal{I}_{k}}^{T}\mathbf{F}_{k, \mathcal{I}_{k}}]_{i,j} 
&=& \sum_{h=1}^{p} I(h \in \mathcal{I}_{k}) F_{hi}^{k} F_{hj}^{k}. \nonumber 
\end{eqnarray}
Similarly, the $i^{th}$ component of the vector $\mathbf{F}_{k}^{T}\mathbf{f}_{k} \in \mathbb{R}^{m_{k}}$,
where $\mathbf{f}_{k} = (f_{1}^{k}, \ldots, f_{p}^{k})^{T}$, is
\begin{eqnarray}
[\mathbf{F}_{k}^{T}\mathbf{f}_{k}]_{i} &=& \sum_{h=1}^{p} F_{hi}^{k} f_{h}^{k} \nonumber \\
&=&  \sum_{h=1}^{p} I(h \in \mathcal{I}_{k}) F_{hi}^{k} f_{h}^{k} + 
\sum_{h=1}^{p} I(h \in \mathcal{I}_{k}^{c}) F_{hi}^{k} f_{h}^{k} \nonumber \\
&=& \sum_{h=1}^{p} I(h \in \mathcal{I}_{k}) F_{hi}^{k} f_{h}^{k}
= \mathbf{F}_{k,\mathcal{I}_{k}}^{T}\mathbf{f}_{k, \mathcal{I}_{k}}. \nonumber 
\end{eqnarray}

\bigskip

Proposition S1 suggests we can ignore the indices corresponding to all-zero rows in $\mathbf{F}_{k}$ when computing $\bgamma^{(k)}$ in the Anderson acceleration update. This is also true when using the DAAREM extension of AA, because
the subsetted version of the matrix $\mathbf{F}_{k}^{T}\mathbf{F}_{k} + \lambda_{k}\mathbf{I}_{m_{k}}$ will also be unchanged by dropping the rows not in $\mathcal{I}_{k}$. In practice, we usually need to specify a small threshold $\varepsilon_{T}$ for which a row sum of the absolute values of $\mathbf{F}_{k}$ less than this threshold determines whether this row index is ignored in subsequent computations. In our implementation of subsetted DAAREM and NIDAAREM, we only consider the set of indices $\mathcal{I}_{k}( \varepsilon_{T} ) = \{ i \in \{1, \ldots, p\}: \sum_{j=1}^{m_{k}} |F_{ij}^{k}| > \varepsilon_{T}\}$ as the set of indices where the absolute row sum of $\mathbf{F}_{k}$ exceeds the threshold $\varepsilon_{T} = 0.001 \times \big( p^{-1}\sum_{i=1}^{p} \sum_{j=1}^{m_{k}} |F_{ij}^{k}| \big)$, and hence the subsetted Anderson acceleration
extrapolation will not be exactly the same as the original AA update.
In the DAAREM extension of AA, the proximity of the subsetted AA extrapolation to the original AA extrapolation will depend largely on the chosen threshold $\varepsilon_{T}$ and the condition number of the matrix $\mathbf{F}_{k}^{T}\mathbf{F}_{k} + \lambda_{k}\mathbf{I}_{m_{k}}$. The check of the condition number in line 5 of the DAAREM algorithm (Algorithm 3 in the main manuscript) ensures that the condition number does not grow too large, and we have observed, in our experiments so far, that the choice of threshold $\varepsilon_{T}$ ensures that the subsetted DAAREM updates are typically close to the DAAREM updates without subsetting. 

One can also note that, when computing $(\mathbf{X}_{k} + \mathbf{F}_{k})(\mathbf{F}_{k,\mathcal{I}_{k}}^{T}\mathbf{F}_{k,\mathcal{I}_{k}})^{-1}\mathbf{F}_{k,\mathcal{I}_{k}}^{T}\mathbf{f}_{k, \mathcal{I}_{k}}$ in the subsetted Anderson acceleration update, we ignore the rows of $\mathbf{F}_{k}$ not
in $\mathcal{I}_{k}(\varepsilon_{T})$ when calculating the matrix sum $\mathbf{X}_{k} + \mathbf{F}_{k}$ because 
all elements of those rows will are very close to $0$.

\section{Additional Simulation Results} 

The results of additional simulations referred to in the main manuscript are reported in Tables S1 - S7.

\begin{table}[ht]
\centering
\begin{tabular}{l|l rrr rrr}
\toprule
\multirow{2}{*}{$\rho$} & \multirow{2}{*}{Method} & \multicolumn{3}{c}{Number of PGD steps} &
  \multicolumn{2}{c}{Timing} &
  \multicolumn{1}{c}{$\varphi(x)$} \\
\cmidrule(r){3-5}\cmidrule(r){6-7}\cmidrule{8-8}
& & mean & median & std. dev. & mean & median & mean \\
\midrule
0.8 & PGD & 735.7 & 759.0 & 68.7 & 1.4 & 1.4 & -11834.300286 \\ 
    & Nesterov & 657.3 & 674.0 & 101.8 & 1.8 & 1.8 & -11834.300286 \\ 
    & Nesterov w restarts & 140.0 & 140.0 & 10.9 & 0.5 & 0.5 & -11834.300286 \\ 
    & DAAREM (fm) & 98.2 & 95.0 & 11.5 & 0.1 & 0.1 & -11834.300286 \\ 
    & DAAREM (am) & 99.2 & 100.0 & 10.5 & 0.1 & 0.1 & -11834.300286 \\ 
    & NIDAAREM (fm) & 70.6 & 70.0 & 4.9 & 0.2 & 0.2 & -11834.300286 \\ 
    & NIDAAREM (am) & 71.4 & 70.0 & 5.9 & 0.2 & 0.2 & -11834.300286 \\ 
    & RNIDAAREM (am) & 68.4 & 68.0 & 5.0 & 0.2 & 0.2 & 66256.635141 \\ 
    & SDAAREM (am) & 98.2 & 95.0 & 11.5 & 0.1 & 0.1 & -11834.300286 \\ 
    & SNIDAAREM (am) & 70.7 & 70.0 & 5.0 & 0.2 & 0.2 & -11834.300286 \\ 
    & SQUAREM & 117.2 & 117.0 & 12.8 & 0.3 & 0.3 & -11834.300286 \\
\midrule
0.95 & PGD & 9068.1 & 9315.0 & 1005.4 & 17.0 & 17.4 & -11044.692442 \\ 
     & Nesterov & 5604.8 & 5861.5 & 1117.2 & 15.0 & 15.7 & -11044.692440 \\ 
     & Nesterov w restarts & 463.4 & 462.0 & 39.0 & 1.6 & 1.6 & -11044.692440 \\ 
     & DAAREM (fm) & 326.7 & 325.0 & 29.8 & 0.1 & 0.1 & -11044.692441 \\ 
     & DAAREM (am) & 325.5 & 325.0 & 34.3 & 0.1 & 0.1 & -11044.692441 \\ 
     & NIDAAREM (fm) & 206.8 & 206.0 & 17.3 & 0.5 & 0.5 & -11044.692441 \\ 
     & NIDAAREM (am) & 212.9 & 208.0 & 22.8 & 0.5 & 0.5 & -11044.692441 \\ 
     & RNIDAAREM (am) & 238.7 & 231.0 & 46.2 & 0.9 & 0.9 & 51167.054416 \\ 
     & SDAAREM (am) & 326.7 & 325.0 & 29.8 & 0.1 & 0.1 & -11044.692441 \\ 
     & SNIDAAREM (am) & 210.5 & 207.5 & 20.6 & 0.5 & 0.5 & -11044.692441 \\ 
     & SQUAREM & 704.7 & 655.5 & 251.2 & 1.6 & 1.4 & -11044.692441 \\ 
\bottomrule
\end{tabular}
\caption{\small{Simulation study for the $\ell_{1}$-penalized regression problem with $n = 1000$, $p = 100$, and $\lambda = 500$. Mean and median 
number of proximal gradient descent (PGD) iterations required to
converge across 50 simulation replications are shown for each method with number of iterations truncated at 500,000 if not
converged within 500,0000 iterations. Mean and median computational times across simulation replications are also shown. Mean values of the objective function at convergence are in the column labeled by $\varphi(x)$. Numbers with smallest mean/median PGD iterations and smallest mean/median times across methods are in bold text. DAAREM (fm) and DAAREM (am) denote the DAAREM method with the fixed and alternating monotonicity monitoring strategies respectively,
and RNIDAAREM denotes the NIDAAREM procedure that monitors progress in the residuals rather than the objective function. } }  
\label{tab:lasso_results_S1}
\end{table}

\begin{table}[ht]
\centering
\begin{tabular}{l|l rrr rrr}
\toprule
\multirow{2}{*}{$\rho$} & \multirow{2}{*}{Method} & \multicolumn{3}{c}{Number of PGD steps} &
  \multicolumn{2}{c}{Timing} &
  \multicolumn{1}{c}{$\varphi(x)$} \\
\cmidrule(r){3-5}\cmidrule(r){6-7}\cmidrule{8-8}
& & mean & median & std. dev. & mean & median & mean \\
\midrule
0.8 & PGD & 1115.2 & 1115.5 & 60.0 & 2.0 & 2.0 & -60474.347165 \\ 
    & Nesterov & 1231.0 & 1233.5 & 57.3 & 3.1 & 3.1 & -60474.347165 \\ 
    & Nesterov w restarts & 223.8 & 221.5 & 27.0 & 0.7 & 0.7 & -60474.347165 \\ 
    & DAAREM (fm) & 154.8 & 150.0 & 23.4 & 0.5 & 0.5 & -60474.347165 \\ 
    & DAAREM (am) & 161.1 & 150.0 & 30.0 & 0.5 & 0.5 & -60474.347165 \\ 
    & NIDAAREM (fm) & 125.3 & 116.0 & 26.4 & 0.7 & 0.6 & -60474.347165 \\ 
    & NIDAAREM (am) & 125.6 & 115.5 & 26.1 & 0.7 & 0.6 & -60474.347165 \\ 
    & RNIDAAREM (am) & 122.3 & 111.0 & 27.5 & 0.4 & 0.3 & 245521.481243 \\ 
    & SDAAREM (am) & 154.8 & 150.0 & 23.4 & 0.5 & 0.5 & -60474.347165 \\ 
    & SNIDAAREM (am) & 125.0 & 117.0 & 26.5 & 0.7 & 0.6 & -60474.347165 \\ 
    & SQUAREM & 151.4 & 153.0 & 11.9 & 0.3 & 0.3 & -60474.347165 \\ 
\midrule
0.95 & PGD & 14436.2 & 14478.5 & 829.5 & 26.2 & 26.3 & -40676.791280 \\ 
     & Nesterov & 12705.6 & 12859.5 & 1200.6 & 31.7 & 32.0 & -40676.791279 \\ 
     & Nesterov w restarts & 702.4 & 691.5 & 60.5 & 2.2 & 2.2 & -40676.791279 \\ 
     & DAAREM (fm) & 438.3 & 440.0 & 34.5 & 0.6 & 0.6 & -40676.791280 \\ 
     & DAAREM (am) & 447.2 & 447.5 & 33.3 & 0.6 & 0.6 & -40676.791280 \\ 
     & NIDAAREM (fm) & 289.8 & 284.5 & 25.6 & 0.9 & 0.9 & -40676.791280 \\ 
     & NIDAAREM (am) & 302.8 & 299.5 & 26.8 & 0.9 & 0.9 & -40676.791280 \\ 
     & RNIDAAREM (am) & 351.4 & 330.5 & 131.7 & 1.3 & 1.2 & 289447.002035 \\ 
     & SDAAREM (am) & 438.3 & 440.0 & 34.5 & 0.6 & 0.6 & -40676.791280 \\ 
     & SNIDAAREM (am) & 292.0 & 283.0 & 31.2 & 0.9 & 0.9 & -40676.791280 \\ 
     & SQUAREM & 940.9 & 943.5 & 78.4 & 1.9 & 1.9 & -40676.791280 \\ 
\bottomrule
\end{tabular}
\caption{\small{Simulation study for the $\ell_{1}$-penalized regression problem with $n = 2000$, $p = 500$, and $\lambda = 500$. The methods used and are as defined in the caption of Table \ref{tab:lasso_results_S1}.} }  
\label{tab:lasso_results_S2}
\end{table}

\begin{table}[ht]
\centering
\begin{tabular}{l|l rrr rrr}
\toprule
\multirow{2}{*}{$\rho$} & \multirow{2}{*}{Method} & \multicolumn{3}{c}{Number of PGD steps} &
  \multicolumn{2}{c}{Timing} &
  \multicolumn{1}{c}{$\varphi(x)$} \\
\cmidrule(r){3-5}\cmidrule(r){6-7}\cmidrule{8-8}
& & mean & median & std. dev. & mean & median & mean \\
\midrule
0.8 & PGD & 913.0 & 912.5 & 39.8 & 1.7 & 1.7 & -6569.305637 \\ 
    & Nesterov & 1050.6 & 1049.5 & 53.0 & 2.8 & 2.8 & -6569.305637 \\ 
    & Nesterov w restarts & 172.4 & 171.5 & 12.8 & 0.6 & 0.6 & -6569.305637 \\ 
    & DAAREM (fm) & 112.8 & 110.0 & 5.9 & 0.1 & 0.1 & -6569.305637 \\ 
    & DAAREM (am) & 111.7 & 110.0 & 5.9 & 0.1 & 0.1 & -6569.305637 \\ 
    & NIDAAREM (fm) & 118.9 & 117.0 & 6.5 & 0.2 & 0.2 & -6569.305637 \\ 
    & NIDAAREM (am) & 117.7 & 117.0 & 5.4 & 0.2 & 0.2 & -6569.305637 \\ 
    & RNIDAAREM (am) & 119.2 & 117.0 & 6.3 & 0.4 & 0.4 & 250125.797984 \\ 
    & SDAAREM (am) & 112.8 & 110.0 & 5.9 & 0.1 & 0.1 & -6569.305637 \\ 
    & SNIDAAREM (am) & 118.1 & 117.0 & 6.1 & 0.2 & 0.2 & -6569.305637 \\ 
    & SQUAREM & 153.5 & 152.0 & 15.0 & 0.3 & 0.3 & -6569.305637 \\ 
\midrule
0.95 & PGD & 10463.8 & 10582.0 & 516.6 & 19.5 & 19.8 & -6228.420884 \\ 
     & Nesterov & 8442.2 & 8290.5 & 715.3 & 22.3 & 21.9 & -6228.420882 \\ 
     & Nesterov w restarts & 568.6 & 571.0 & 51.0 & 1.9 & 2.0 & -6228.420882 \\ 
     & DAAREM (fm) & 515.5 & 515.0 & 60.2 & 0.1 & 0.1 & -6228.420883 \\ 
     & DAAREM (am) & 430.6 & 427.5 & 35.3 & 0.1 & 0.1 & -6228.420883 \\ 
     & NIDAAREM (fm) & 458.0 & 448.0 & 67.3 & 0.5 & 0.5 & -6228.420883 \\ 
     & NIDAAREM (am) & 386.7 & 388.5 & 23.0 & 0.5 & 0.5 & -6228.420883 \\ 
     & RNIDAAREM (am) & 529.4 & 499.5 & 102.1 & 2.6 & 2.4 & 336719.628735 \\ 
     & SDAAREM (am) & 515.5 & 515.0 & 60.2 & 0.1 & 0.1 & -6228.420883 \\ 
     & SNIDAAREM (am) & 454.6 & 442.5 & 92.9 & 0.5 & 0.5 & -6228.420883 \\ 
     & SQUAREM & 1184.4 & 1191.0 & 88.8 & 2.5 & 2.6 & -6228.420883 \\ 
\bottomrule
\end{tabular}
\caption{\small{Simulation study for the $\ell_{1}$-penalized regression problem with $n = 10000$, $p = 100$, and $\lambda = 50$. The methods used and are as defined in the caption of Table \ref{tab:lasso_results_S1}.} }  
\label{tab:lasso_results_S3}
\end{table}

\begin{table}[ht]
\centering
\begin{tabular}{l|l rrr rrr}
\toprule
\multirow{2}{*}{$\rho$} & \multirow{2}{*}{Method} & \multicolumn{3}{c}{Number of PGD steps} &
  \multicolumn{2}{c}{Timing} &
  \multicolumn{1}{c}{$\varphi(x)$} \\
\cmidrule(r){3-5}\cmidrule(r){6-7}\cmidrule{8-8}
& & mean & median & std. dev. & mean & median & mean \\
\midrule
0.8 & PGD & 1350.4 & 1342.5 & 69.0 & 2.4 & 2.4 & -7807.340216 \\ 
     & Nesterov & 1457.6 & 1453.0 & 71.6 & 3.6 & 3.6 & -7807.340216 \\ 
     & Nesterov w restarts & 245.3 & 239.5 & 31.2 & 0.8 & 0.8 & -7807.340216 \\ 
     & DAAREM (fm) & 169.6 & 170.0 & 7.6 & 0.5 & 0.5 & -7807.340216 \\ 
     & DAAREM (am) & 169.9 & 170.0 & 8.0 & 0.5 & 0.5 & -7807.340216 \\ 
     & NIDAAREM (fm) & 153.4 & 147.0 & 20.4 & 0.7 & 0.6 & -7807.340216 \\ 
     & NIDAAREM (am) & 155.9 & 151.0 & 19.9 & 0.7 & 0.6 & -7807.340216 \\ 
     & RNIDAAREM (am) & 156.5 & 151.5 & 19.4 & 0.5 & 0.5 & 456445.077450 \\ 
     & SDAAREM (am) & 169.6 & 170.0 & 7.6 & 0.5 & 0.5 & -7807.340216 \\ 
     & SNIDAAREM (am) & 152.9 & 146.0 & 20.6 & 0.7 & 0.6 & -7807.340216 \\ 
     & SQUAREM & 227.3 & 225.0 & 16.5 & 0.5 & 0.5 & -7807.340216 \\ 
\midrule
0.95 & PGD & 14903.5 & 14840.0 & 699.8 & 26.9 & 26.8 & -5796.475667 \\ 
     & Nesterov & 12715.6 & 12651.5 & 748.1 & 31.6 & 31.4 & -5796.475667 \\ 
     & Nesterov w restarts & 759.2 & 741.5 & 73.6 & 2.4 & 2.4 & -5796.475667 \\ 
     & DAAREM (fm) & 724.2 & 725.0 & 33.3 & 0.8 & 0.8 & -5796.475667 \\ 
     & DAAREM (am) & 681.4 & 680.0 & 33.0 & 0.8 & 0.8 & -5796.475667 \\ 
     & NIDAAREM (fm) & 452.1 & 452.5 & 31.3 & 1.0 & 1.0 & -5796.475667 \\ 
     & NIDAAREM (am) & 430.5 & 427.5 & 31.2 & 1.0 & 1.0 & -5796.475667 \\ 
     & RNIDAAREM (am) & 822.9 & 796.5 & 175.8 & 4.1 & 3.9 & 228136.157254 \\ 
     & SDAAREM (am) & 724.2 & 725.0 & 33.3 & 0.8 & 0.8 & -5796.475667 \\ 
     & SNIDAAREM (am) & 462.7 & 462.0 & 32.4 & 1.0 & 1.0 & -5796.475667 \\ 
     & SQUAREM & 1431.5 & 1425.0 & 73.0 & 2.9 & 2.9 & -5796.475667 \\ 
\bottomrule
\end{tabular}
\caption{\small{Simulation study for the $\ell_{1}$-penalized regression problem with $n = 2000$, $p = 500$, and $\lambda = 50$. The methods used and are as defined in the caption of Table \ref{tab:lasso_results_S1}.} }  
\label{tab:lasso_results_S4}
\end{table}

\begin{table}[ht]
\centering
\begin{tabular}{l|l rrr rrr}
\toprule
\multirow{2}{*}{$\rho$} & \multirow{2}{*}{Method} & \multicolumn{3}{c}{Number of PGD steps} &
  \multicolumn{2}{c}{Timing} &
  \multicolumn{1}{c}{$\varphi(x)$} \\
\cmidrule(r){3-5}\cmidrule(r){6-7}\cmidrule{8-8}
& & mean & median & std. dev. & mean & median & mean \\
\midrule
0.8 & PGD & 489760.8 & 500000.0 & 33747.6 & 874.6 & 890.2 & -12657.185676 \\ 
    & Nesterov & 163376.5 & 164762.0 & 43460.1 & 415.1 & 419.7 & -12657.158472 \\ 
    & Nesterov w restarts & 6681.0 & 6326.0 & 2904.9 & 21.9 & 20.7 & -12657.158472 \\ 
    & DAAREM (fm) & 5797.5 & 3872.5 & 8824.6 & 32.9 & 21.9 & -12657.158503 \\ 
    & DAAREM (am) & 4784.6 & 4655.0 & 1286.3 & 32.9 & 21.9 & -12657.158503 \\ 
    & NIDAAREM (fm) & 8743.0 & 5072.5 & 19111.6 & 50.2 & 30.1 & -12657.158504 \\ 
    & NIDAAREM (am) & 4437.1 & 4396.0 & 1373.9 & 25.5 & 25.1 & -12657.158502 \\ 
    & RNIDAAREM (am) & 23429.1 & 11942.5 & 69126.7 & 164.7 & 82.4 & -12657.158512 \\ 
    & SDAAREM (am) & 4824.4 & 3770.0 & 6167.0 & 14.6 & 11.5 & -12657.158504 \\ 
    & SNIDAAREM (am) & 3388.5 & 2379.0 & 6298.0 & 10.1 & 7.1 & -12657.158502 \\ 
    & SQUAREM & 8055.8 & 7191.5 & 4104.9 & 17.8 & 15.8 & -12657.158501 \\ 
\midrule
0.95 &  PGD & 500000.0 & 500000.0 & 0.0 & 890.3 & 888.7 & -14331.365228 \\ 
     &  Nesterov & 302468.9 & 292472.5 & 77602.6 & 765.9 & 742.0 & -14331.168597 \\ 
     &  Nesterov w restarts & 10516.2 & 9644.5 & 4509.8 & 34.3 & 31.2 & -14331.168599 \\ 
     &  DAAREM (fm) & 8312.5 & 5495.0 & 11869.8 & 46.6 & 30.8 & -14331.168653 \\ 
     &  DAAREM (am) & 7699.1 & 6032.5 & 7309.4 & 46.6 & 30.8 & -14331.168653 \\ 
     &  NIDAAREM (fm) & 8192.6 & 6397.0 & 8719.1 & 47.1 & 37.0 & -14331.168662 \\ 
     &  NIDAAREM (am) & 6714.7 & 6209.5 & 2046.2 & 38.9 & 35.9 & -14331.168658 \\ 
     &  RNIDAAREM (am) & 31634.4 & 19465.0 & 68462.2 & 221.6 & 133.8 & -14331.168771 \\ 
     &  SDAAREM (am) & 7704.2 & 5465.0 & 10247.5 & 23.1 & 16.5 & -14331.168654 \\ 
     &  SNIDAAREM (am) & 5919.7 & 3784.5 & 10347.1 & 17.5 & 11.2 & -14331.168652 \\ 
     &  SQUAREM & 24663.5 & 11514.5 & 69790.9 & 55.8 & 25.2 & -14331.168662 \\ 
\bottomrule
\end{tabular}
\caption{\small{Simulation study for the $\ell_{1}$-penalized regression problem with $n = 100$, $p = 10000$, and $\lambda = 50$. The methods used and are as defined in the caption of Table \ref{tab:lasso_results_S1}.} }  
\label{tab:lasso_results_S5}
\end{table}

\begin{table}[ht]
\centering
\begin{tabular}{lrr}
\toprule
Method & Number of PGD steps  & Time (in seconds) \\ 
\midrule
PGD & 56035 & 221.1 \\ 
  Nesterov & 23719 & 129.4 \\ 
  Nesterov (restart) & 2445 & 17.0 \\ 
  SQUAREM & 2509 & 11.2 \\ 
  DAAREM (fm) & 585 & 3.5 \\ 
  DAAREM (am) & 535 & 3.2 \\ 
  NIDAAREM (fm) & 882 & 5.7 \\ 
  NIDAAREM (am) & 782 & 5.2 \\ 
  MPE & 2516 & 10.7 \\ 
   \bottomrule
\end{tabular}
\caption{{\small Results for $\ell_{1}$-penalized logistic regression using the \textbf{Madelon} data with warm starts over a sequence of $10$ penalty terms. The total number of PGD steps required to compute the entire sequence of $\ell_{1}$-penalized logistic regression solutions is shown for each method. In addition, the time
required to compute the entire solution path is shown. }} 
\label{tab:madelon_warm}
\end{table}

\begin{table}[ht]
\centering
\begin{tabular}{l|l rrr rrr}
\toprule
\multirow{2}{*}{$\lambda$} & \multirow{2}{*}{Method} & \multicolumn{3}{c}{Number of PGD steps} &
  \multicolumn{2}{c}{Timing} &
  \multicolumn{1}{c}{$\varphi(x)$} \\
\cmidrule(r){3-5}\cmidrule(r){6-7}\cmidrule{8-8}
& & mean & median & std. dev. & mean & median & mean \\
\midrule
200 & PGD & 167.0 & 167.0 & 0.0 & 686.3 & 688.0 & -455145.740802 \\ 
    & Nesterov & 235.0 & 235.0 & 0.0 & 1652.7 & 1626.6 & -455145.740802 \\ 
    & Nesterov (restart) & 89.4 & 89.0 & 1.8 & 890.7 & 870.2 & -455145.740802 \\ 
    & SQUAREM & 108.0 & 108.0 & 0.0 & 630.9 & 625.6 & 455145.740802 \\ 
    & DAAREM (am) & 42.5 & 42.5 & 2.6 & 337.3 & 329.0 & -455145.740802 \\ 
    & NIDAAREM (am) & 32.0 & 30.0 & 2.6 & 402.4 & 376.7 & -455145.740802 \\ 
    & RNIDAAREM (resid-am) & 32.0 & 30.0 & 2.6 & 399.0 & 378.4 & -455145.740802 \\ 
    & SoftImpute &  NA &  NA &  NA & 719.6 & 706.1 & -455145.740804 \\
\bottomrule
\end{tabular}
\caption{Results for the matrix completion simulation study using the movielens data. This table shows performance results for the case where $10$ different random starting values were used and where the penalty term $\lambda$ was set to $\lambda = 200$.} 
\label{tab:movielens_results}
\end{table}

\end{document}